\documentclass[fleqn,usenatbib]{mnras}
\usepackage{newtxtext,newtxmath}

\usepackage[T1]{fontenc}

\DeclareRobustCommand{\VAN}[3]{#2}
\let\VANthebibliography\thebibliography
\def\thebibliography{\DeclareRobustCommand{\VAN}[3]{##3}\VANthebibliography}

\usepackage{array}
\usepackage{subfigure}
\usepackage{graphicx}	
\usepackage{amsmath}	
\usepackage{multicol}   
\usepackage{bm}		    
\usepackage{pdflscape}	
\usepackage{soul}

\usepackage{color}

\usepackage{xcolor}

\newcommand{\Mpc}{\ensuremath{\,{\rm Mpc}}}

\newcommand{\Msun}{\ensuremath{\, {\rm M}_{\odot}}}

\newcommand{\hi}{H{\textbf{\sc\ i}}}

\usepackage[T1]{fontenc}
\usepackage{ae,aecompl}

\usepackage{newtxtext,newtxmath}

\title[\hi\ Content vs Group-Centric Radius]
{The Atomic Hydrogen Content of Galaxies as a function of Group-Centric Radius}

\author[Wenkai Hu et al.]
{Wenkai Hu$^{1,2,3,6}$\thanks{Contact e-mail: \href{wenkai.hu@lam.fr}{wenkai.hu@lam.fr}},
Luca Cortese$^{1,3}$\thanks{Contact e-mail: \href{luca.cortese@uwa.edu.au}{luca.cortese@uwa.edu.au}},
Lister Staveley-Smith$^{1,3}$, 
Barbara Catinella$^{1,3}$,
\newauthor
Garima Chauhan$^{1,3}$,
Claudia del P. Lagos$^{1,3}$,
Tom Oosterloo$^{4,5}$,
Xuelei Chen$^{6,7,8}$
\\
$^{1}$ International Centre for Radio Astronomy Research (ICRAR), M468, University of Western Australia, 35 Stirling Hwy, WA 6009, Australia\\
$^{2}$ Aix Marseille Universit$\rm\acute{e}$, CNRS, LAM (Laboratoire d$^{\ \prime}$Astrophysique de Marseille), F-13388 Marseille, France\\
$^{3}$ ARC Centre of Excellence for All Sky Astrophysics in 3 Dimensions (ASTRO 3D), Australia\\
$^{4}$ ASTRON, the Netherlands Institute for Radio Astronomy, Postbus 2, 7990 AA, Dwingeloo, The Netherlands\\
$^{5}$ Kapteyn Astronomical Institute, University of Groningen, P.O. Box 800, 9700 AV Groningen, The Netherlands\\
$^{6}$ National Astronomical Observatories, Chinese Academy of Sciences, 20A, Datun Road, Chaoyang District, Beijing 100101, China\\
$^{7}$ Center of High Energy Physics, Peking University, Beijing 100871, China\\
$^{8}$ School of Astronomy and Space Science, University of Chinese Academy of Sciences, Beijing 100049, China
}

\date{Last updated 2020 May 22; in original form 2019 September 5}

\pubyear{2021}

\begin{document}
\label{firstpage}
\pagerange{\pageref{firstpage}--\pageref{lastpage}}
\maketitle

\setstcolor{red}

\begin{abstract}

We apply a spectral stacking technique to Westerbork Synthesis Radio Telescope observations to measure the neutral atomic hydrogen content (\hi) of nearby galaxies in and around galaxy groups at $z < 0.11$. Our sample includes 577 optically-selected galaxies (120 isolated galaxies and 457 satellites) covering stellar masses between 10$^{10}$ and 10$^{11.5}$ M$_{\odot}$, cross-matched with Yang's group catalogue, with angular and redshift positions from the Sloan Digital Sky Survey. We find that the satellites in the centres of groups have lower \hi\ masses at fixed stellar mass and morphology (characterised by the inverse concentration index) relative to those at larger radii. These trends persist for satellites in both high-mass ($M_{\rm halo} > 10^{13.5}h^{-1}$M$_{\odot}$) and low-mass ($M_{\rm halo} \leqslant 10^{13.5}h^{-1}$M$_{\odot}$) groups, but disappear if we only consider group members in low local density ($\Sigma <$ 5 gal/Mpc$^{-2}$) environments. Similar trends are found for the specific star formation rate. Interestingly, we find that the radial trends of decreasing \hi\ mass with decreasing group-centric radius extend beyond the group virial radius, as isolated galaxies close to larger groups lack \hi\ compared with those located more than $\sim$3.0 $R_{180}$ away from the center of their nearest group. We also measure these trends in the late-type subsample and obtain similar results. Our results suggest that the \hi\ reservoir of galaxies can be affected before galaxies become group satellites, indicating the existence of pre-processing in the infalling isolated galaxies. 
\end{abstract}





\vspace*{-0.5cm}
\section{Introduction}

It is widely accepted that the evolution of a galaxy is significantly influenced by its environment \citep{1984ARA&A..22..185D,2009ARA&A..47..159B,2010PhR...495...33B}. Galaxies can be depleted in \hi\ by (i) directly removing the cold gas via interaction with the intra-cluster medium (ram-pressure stripping), or with the parent halo, or with other galaxies (tidal interaction, harassment); and (ii) by reducing the rate at which the galaxies accrete the gas from their halos (strangulation). 

A number of observational studies show that galaxies in dense regions are redder and have lower star formation rate than those in the field \citep{1983AJ.....88..483K,1999ApJ...527...54B,1999ApJ...518..576P,2002MNRAS.334..673L,2003ApJ...584..210G,2004MNRAS.353..713K,2009MNRAS.393.1324B,2010ApJ...721..193P,2010MNRAS.409..337C,2012ApJ...757....4P,2017MNRAS.464..121S}, with galaxies in dense regions invariably showing \hi\ deficiency \citep{1973MNRAS.165..231D,1984ARA&A..22..445H,1985ApJ...292..404G,2001ApJ...548...97S,2011MNRAS.415.1797C,2013MNRAS.436...34C,2016ApJ...824..110O,2016ApJ...832..126S,2017MNRAS.466.1275B,2020MNRAS.493.5024L,2021arXiv210402193C}, and gas-rich galaxies most often being found in weakly clustered regions \citep{2007ApJ...654..702M,2012ApJ...750...38M}.

Since galaxy evolutionary pathways are affected by gravitational and hydrodynamical interactions with halos, the position of a galaxy relative to the centre of its halo is important. In practice, the centre is often defined as the position of the most massive galaxy or the luminosity-weighted centre for all group members and, in recent years, significant effort has been invested into exploring the correlation between galaxy properties and group-centric radius. Using galaxy group catalogues based on the Sloan Digital Sky Survey \citep[SDSS;][]{2000AJ....120.1579Y} Data Release 7 \citep[DR7;][]{2009ApJS..182..543A}, \citet{2012MNRAS.424..232W} conclude that the quenched galaxy fraction increases at smaller group-centric radii. \citet{2016MNRAS.462.2559B} quantify the impact that various environmental and galactic properties have on the quenching of star formation, using a sample of $\sim 400,000$ centrals and $\sim 100,000$ satellites from SDSS 7. They find that the group-centric radius has a significant impact on the quenched fraction at fixed central velocity dispersion, with smaller group-centric radius being associated with a higher quenched fraction. \citet{2018ApJ...860..102W} show that the correlation between the quenched fraction of centrals and the halo-centric radius is weak, while the quenched fraction of satellites, in a given halo mass bin, shows a decreasing trend from the group centre outward. However, when both stellar and halo mass are controlled, the quenched fractions of centrals and satellites are once again similar (albeit some differences remain; see \citealt{2019MNRAS.483.5444D} and discussion therein).

As well as the information derived from optical bands, the knowledge of a galaxy's \hi\ content, as the raw fuel for future star formation, is also critical in understanding the influence of environment on galaxy evolution. Observations of \hi\ gas have the potential to reveal the removal of cold gas. However, it is difficult for the current generation of radio telescopes to directly measure the 21-cm emission of large galaxy samples across environments, so this approach is limited to low redshifts, and generally able to trace only gas-normal galaxies. Using the method of spectral stacking, we can push the measurement of \hi\ gas content to higher redshifts. The technique combines a large number of rest-frame spectra extracted from the radio data with angular and redshift positions from optical catalogues. In this process, the noise is averaged down and a more significant, but averaged over a large sample of galaxies, spectral-line signal is recovered. By stacking a larger number of galaxies, considerable large effective survey volumes and much smaller cosmic variance can be obtained. These advantages make stacking a powerful and easy-to-use tool for statistical studies of \hi\ properties as a function of environment (e.g., \citet{2012MNRAS.427.2841F,2017MNRAS.466.1275B}). 

\citet{2016ApJ...824..110O} used stacking and multiple linear regressions, presenting the distribution of \hi\ content in nearby groups and clusters measured in the 70$\%$ complete Arecibo Legacy Fast-ALFA \citep[ALFALFA;][]{2005AJ....130.2598G} survey. They find that at fixed stellar mass, the late-type galaxies in the inner regions of groups lack \hi\ compared with galaxies in a control region extending to 4.0 Mpc surrounding each group. 

In this work, we use a stacking method to explore how the \hi\ content and the specific star formation rate (sSFR) change with the group-centric radius. In \citet[][hereafter Paper I]{10.1093/mnras/stz2038} and \citet{2020MNRAS.493.1587H}, we have developed an interferometric stacking technique to measure the \hi\ content in galaxies and confirm that there is little evolution in cosmic \hi\ density ($\rm \Omega_{\hi}$) at low redshift ($z$ < 0.11). This paper uses the same sample and technique to study the relation between \hi\ content and group-centric radius.

This paper is organized as follows: Section~\ref{sec:sample} presents the observational data and the optical catalogue we used in this paper. In Section~\ref{sec:script}, we give a simple description of the stacking methodology. In Section~\ref{sec:results}, we measure the \hi\ content in satellites with different group-centric radii. We discuss our results in Section~\ref{sec:discussion} and give the summary
in Section~\ref{sec:summary}. Throughout this paper, we use the cosmological parameters $h=0.7$, $\Omega_{\rm m} = 0.3$ and $\Omega_{\Lambda} = 0.7$.

\vspace*{-0.5cm}
\section{Sample}
\label{sec:sample}
\subsection{Radio Data}
The \hi\ observations were conducted with the Westerbork Synthesis Radio Telescope (WSRT) in a strip in the SDSS South Galactic Cap (21h < RA < 2h and $10^{\circ}$< Dec <$16^{\circ}$), during May 2011 to October 2012. The data are fully described in Paper I. We used 351 hours of telescope time to observe 36 individual pointings, with each pointing being observed for between 5 hr and 12 hr. Due to bad data quality, data from one of the pointings were discarded. The half-power beam width (HPBW) of WSRT is 35 arcmins, and the average synthesized beam size is $108\arcsec \times 22\arcsec$. The frequency for the reduced data ranges from 1.406 GHz to 1.268 GHz, corresponding to a redshift range of $0.01 < z < 0.12$. The radio astronomy data reduction package {\sc miriad} \citep{1995ASPC...77..433S} was used to reduce and calibrate the data. The reduced data cubes of each pointing have a size of $1^{\circ}\times1^{\circ}$ with resolution of $3\arcsec \times 3\arcsec$ in pixel size and 0.15625 MHz in frequency channel.

\subsection{SDSS}
We use SDSS 
DR7 as the optical catalogue for our stacking analysis. In our cross-matching procedure, we first obtain the positions (redshift,ra,dec) of the optical galaxies, then we extract the spectra in the same positions in the radio data. By cross-matching the SDSS catalogue with our radio data, we obtain a sample of 1895 galaxies with a redshift range of $0.01 < z < 0.11$ and an average redshift of $\langle z\rangle$ = 0.066.

\begin{figure}
    \centering
    \includegraphics[width=8cm]{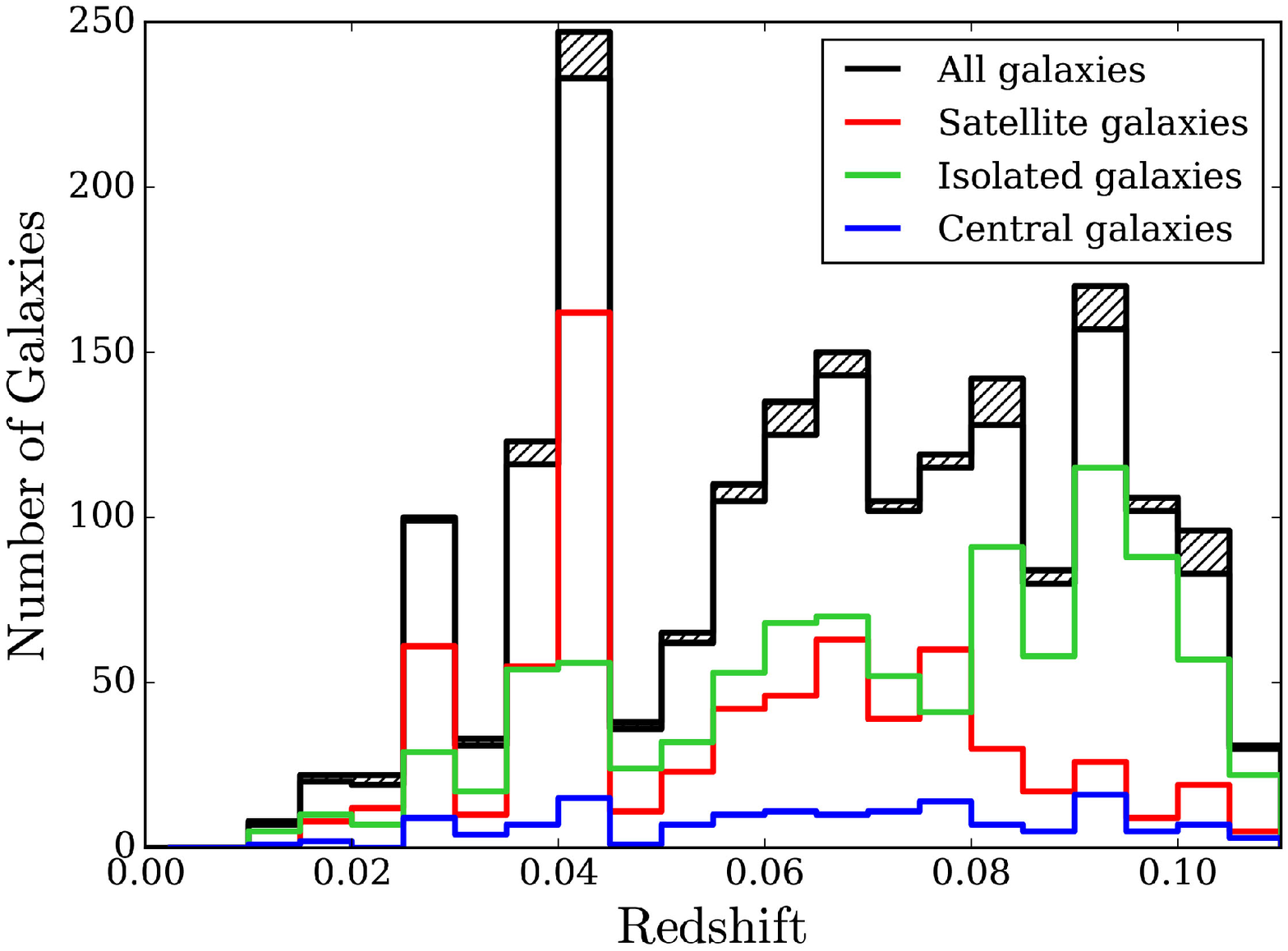}
    \includegraphics[width=8cm]{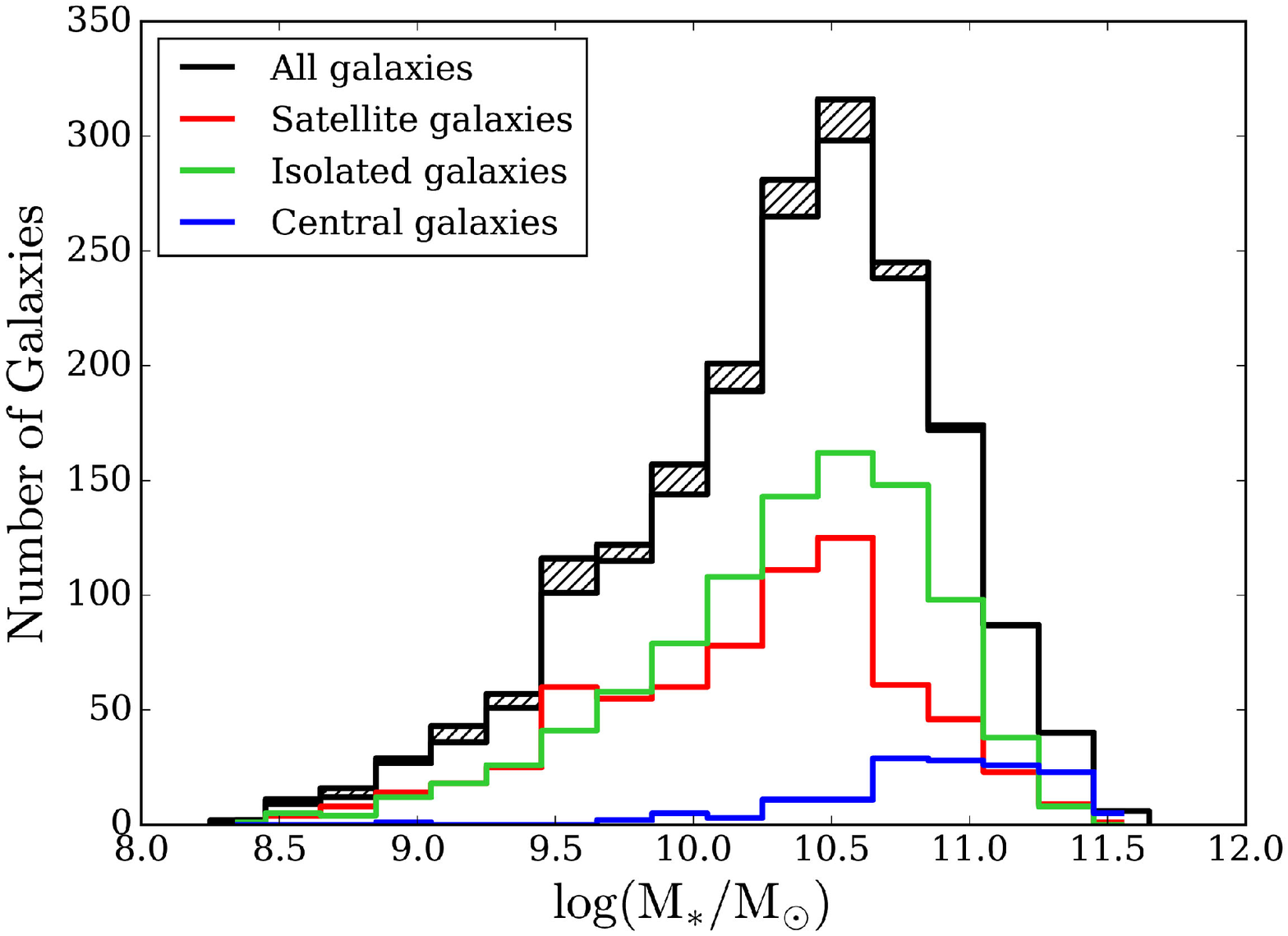}
    \caption{Redshift (top panel) and stellar mass (bottom panel) distribution for the SDSS spectroscopic sub-sample contained within the 35 WSRT pointings. The distributions for all galaxies, satellite galaxies, isolated and group central galaxies are shown by the black, red, green and blue lines, respectively. The hatched region shows the galaxies not included in Yang catalogue. The group catalogue is based on the main galaxy sample from SDSS DR7 and galaxies with a redshift completeness C $\leqslant$ 0.7 are removed. The missing galaxies tend to be low-mass and faint. The interval for stellar mass bins and redshift bins are 0.2 dex and 0.005, respectively. The lower limit of the redshift of the selected sample is $z = 0.01$, and the upper limit of the redshift is $z = 0.11$. The average redshift of the sample is $\langle z\rangle = 0.065$.}
    \label{distribution}
\end{figure}

\begin{figure}
    \centering
    \includegraphics[width=8cm]{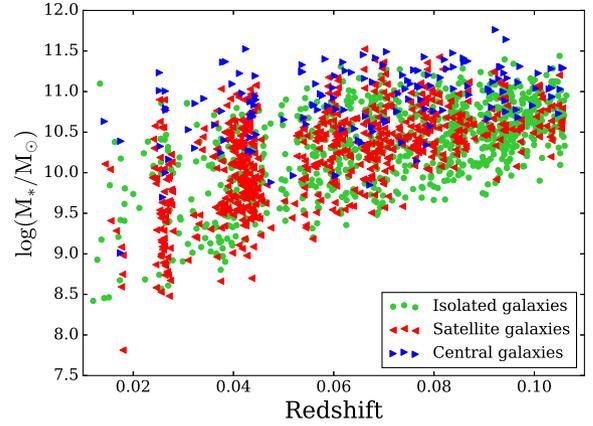}
    \caption{The stellar mass vs. redshift distribution for the SDSS spectroscopic sub-sample contained within the 35 WSRT pointings. The satellite, group centrals and isolated galaxies are showed by the red left triangles, blue right triangles and green points, respectively.}
    \label{stellarmass_redshift}
\end{figure}

\subsection{Yang Group Catalogue}
In this paper, we use a dark matter halo group catalogue \citep{2007ApJ...671..153Y,2012ApJ...752...41Y} based on the galaxies in the SDSS main galaxy sample with redshift completeness C $\geq$ 0.7. The first Yang group catalogue derived from the SDSS DR4 \citep{2007ApJ...671..153Y} used 362,356 galaxies to identify groups in the redshift range $0.01 < z < 0.2$. Extending their analysis to SDSS DR7, the number of galaxies was increased to $\sim 599,300$ \citep{2012ApJ...752...41Y}.
In this catalogue, the following iterative process is used to identify the dark matter halos: (1) identification of the group centres; (2) calculation of the group luminosity; and (3) estimation of additional properties of the tentative group such as mass, size and velocity dispersion. Based on the properties of the halos, the group members might be reassigned. Afterwards, a new group centre is then defined and the process is iterated until the members of the groups are stable. The final halo masses are calculated using the halo mass function derived by \citet{2006ApJ...646..881W}.

In this paper, the central galaxy is determined as the galaxy with highest stellar mass. Other galaxies in the group will be called satellites. The stellar masses and sSFR used in this paper are taken from the MPA-JHU (Max-Planck Institute for Astrophysics - John Hopkins University) value-added galaxy catalogue, which are derived from the SDSS DR7 catalogue \citep{2003MNRAS.341...33K}.

\begin{figure}
    \centering
    \includegraphics[width=8.0cm]{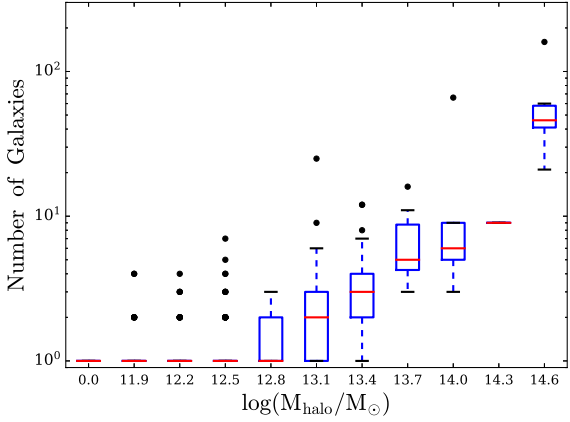}
    \caption{Box-and-whisker plot of number of galaxies in a halo of a certain mass. The width of halo mass bins is 0.3 in log scale. The red line indicates the mean number of galaxies in a halo and the box marks the first and third quartile of the data, while the whiskers mark the lowest and highest value still within 1.5 times the interquartile range. The outliers are marked as individual black dots. The figure shows that the median number of galaxies per halo increases with increasing halo mass and that the range in number of galaxies gets broader. Note that $\log M_{\rm halo} = 0$ means that no mass was assigned to that halo.}
    \label{boxplot}
\end{figure}

In this group catalogue, galaxies with a redshift completeness C $\leqslant$ 0.7 are excluded. The cross-matching of the SDSS DR7 sources contained within our pointings with the Yang catalogue (DR7) reduces the number of galaxies in the sample by $\sim 5\%$ from 1895 to 1793 galaxies, with an average redshift of $\langle z\rangle$ = 0.065. The redshift and stellar mass distribution of satellites (red line), isolated galaxies (green line), centrals (blue line) and all galaxies (black line) are shown in Figure~\ref{distribution}. Most of the galaxies in our sample have a stellar mass of around 10$^{10.5}$M$_{\odot}$. In Figure~\ref{stellarmass_redshift}, we present the stellar mass distribution against redshift. This shows that our sample is magnitude-limited. We show the number of galaxies in bins of halo mass in Figure~\ref{boxplot}. It shows that the median number of galaxies per halo increases with increasing halo mass. Groups with higher halo mass have more satellite galaxies.

Out of the 1793 galaxies included in the group catalog, 350 do not have assigned halo masses as the Yang catalogue did not assign halo masses to small halos. Among these galaxies, over 96 percent are isolated centrals. For these galaxies in extremely small halo mass groups,  we estimate their halo masses by solving the following equation \citep{2009ApJ...693..830Y}:
\begin{eqnarray}
     \langle M_{\ast}\rangle(M_{h}) = M_{0} \frac{(M_{h}/M_{1})^{\alpha+\beta}}{(1+M_{h}/M_{1})^{\beta}},
    \label{halomass_estimation}
\end{eqnarray}
where $M_{\ast}$ is the stellar mass of the central galaxies, $M_{h}$ is the halo mass, $M_{1}$ is a characteristic halo mass. The parameters obtained from the SDSS groups are: $\alpha = 0.315$, $\beta = 4.543$, $\log M_{0} = 10.306$, and $\log M_{1} = 11.040$, where $M_{0}$ is in units of $h^{-2}$ M$_{\odot}$ and $M_{1}$ in $h^{-1}$ M$_{\odot}$.

\begin{figure}
    \centering
    \includegraphics[width=8cm]{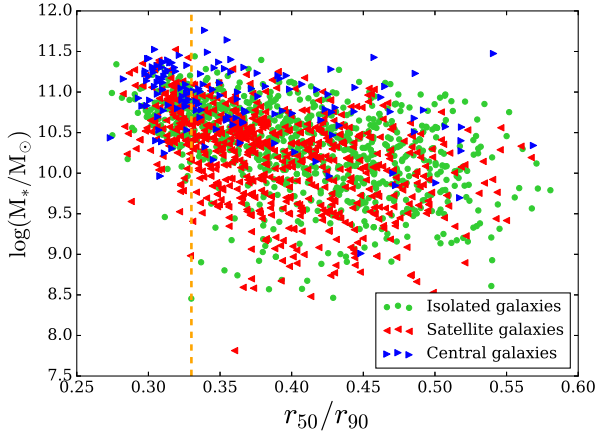}
    \caption{Same as Figure~\ref{stellarmass_redshift}, but for the stellar mass vs. inverse concentration index distribution. The orange-dashed line shows the morphological selection we use (inverse concentration index $r_{50}/r_{90}$ = 0.33).}
    \label{stellarmass_concentration}
\end{figure}

Across the sample, 699 (39$\%$) galaxies are classified as satellite galaxies while 1094 (61$\%$) galaxies are classified as central galaxies, among which 906 galaxies are isolated. In the following calculation, we will consider the central galaxies and isolated galaxies separately. In order to make sure that we are tracing environment instead of simple selection biases, we account for galaxy type and stellar mass by limiting the mass range to $10.0 < \log (M_{\ast}/$M$_{\odot}) < 11.5$ when comparing \hi\ properties in the inner and outer regions of groups. We also create a sub-sample that only contains the late-type galaxies to quantify the influence from morphology. We separate galaxies by the morphology using the r-band inverse concentration index, which is defined as the ratio of the radii containing 50$\%$ ($r_{50}$) to that containing 90$\%$ ($r_{90}$) of the Petrosian flux. Galaxies with $r_{50}/r_{90}$ larger than 0.33 are classified as late type \citep{2001AJ....122.1238S}. In Figure~\ref{stellarmass_concentration}, we present the stellar mass distribution against inverse concentration index. As expected, early-type galaxies tend to have larger stellar mass. The orange-dashed line shows the morphological selection we used.  The mass-limited subsample contains 466 satellite galaxies and 723 isolated galaxies, among which 457 (344 for the late-type subsample) satellites with group-centric radius $R < R_{180}$ and 120 (90 for the late-type subsample) isolated galaxies in the infalling region (see the definition in Section~\ref{sec:larger_distance}) are used in the following to quantify the \hi\ content of galaxies as a function of group-centric radius.

Limited by the size of the WSRT primary beam, some groups have not been mapped out to their virial radius. Specifically, out of a total of 236 groups (120 isolated galaxies included) in the mass limited sample, 23 groups are partially observed. This will limit our \hi\ measurement at large group-centric radius, but the relationship between \hi\ properties and group-centric radius should not be affected.

\vspace*{-0.5cm}
\section{Stacking Script}
\label{sec:script}
\subsection{Main Procedure}
The stacking technique applied in this work is described in detail in Paper I. Here, we provide a short summary of the key steps. The spectra for galaxies were extracted from the radio data cubes over an extended region with aperture radius of 35kpc around the SDSS position. After removing the residual continuum emission from bright sources, the spectra were de-redshifted to rest-frame and the \hi\ flux density is conserved via using $S_{\nu_{\rm res}} = S_{\nu_{\rm obs}}/(1+z)$. The mass spectra were calculated using:
\begin{eqnarray}
    m_{\rm \hi}(\nu) = 4.98\times10^{7} S_{\nu} D_{L}^{2}f^{-1} ,
    \label{mass_spectrum}
\end{eqnarray}
where $S_{\nu}$ is the de-redshifted \hi\ flux density in Jy, $D_{L}$ refers to the luminosity distance in units of Mpc, $f$ denotes the normalised primary beam response, and $m_{\rm \hi}$ is in units of M$_{\sun}$ MHz$^{-1}$. The spectrum of $i$-th galaxy is weighted by:
\begin{eqnarray}
    w_{i} = f^{2}D_{L}^{-1}\sigma^{-2},
    \label{weight}
\end{eqnarray}
where 
$\sigma$ is the rms noise of the flux density spectra. The final averaged stacked spectrum is calculated by:
\begin{eqnarray}
    \langle m_{\rm \hi}(\nu)\rangle = \frac{\sum_{i=1}^{n}w_{i}m_{\rm \hi,i}}{\sum_{i=1}^{n}w_{i}}.
    \label{weighted_spectrum}
\end{eqnarray}
The integrated \hi\ mass of a stack can be obtained by integrating the mass spectrum along the frequency axis:
\begin{eqnarray}
    \langle M_{\rm \hi}\rangle = \int\langle m_{\rm \hi}(\nu)\rangle d\nu,
    \label{integrated_mass}
\end{eqnarray}
where the integration region is $\sim \pm1.5$ MHz and will be slightly adjusted to exclude external noise and signal.


The error of the \hi\ mass measurement is estimated by jackknife resampling. 5 per cent of the sample are excluded when we do the jackknife resampling.
The \hi\ mass fraction $\langle M_{\rm \hi}/M_{\ast}\rangle$ and its error can also be measured by stacking the individual $M_{\rm \hi}/M_{\ast}$ spectra. This is done via Equation~\ref{weighted_spectrum} and \ref{integrated_mass}, with $M_{\rm \hi}$ replaced by $M_{\rm \hi}/M_{\ast}$.

\subsection{Confusion Correction}
The value of the average \hi\ mass we measure via the stacking method may be artificially inflated by the effect of beam confusion. Flux from objects with a projected distance smaller than the WSRT synthesised beam and line-of-sight velocity difference lower than the spectral extraction region cannot be distinguished. Additional \hi\ signal from nearby galaxies can therefore contaminate the measurements. We define confused galaxies as SDSS galaxies having one or more companions within the WSRT synthesized beam, and within 3 MHz ($\sim$ 630 km s$^{-1}$ at $z=0$) in frequency. No stellar mass cut is applied when we select the confused galaxies. Although the WSRT synthesized beam is small, $\sim 7$ per cent of our whole sample is potentially confused with neighbouring galaxies. Directly removing confused galaxies will have the effect of removing massive centrals and gas-rich satellites. To correct for the effects of confusion in these galaxies without introducing bias, we follow the method in \citet{2012MNRAS.427.2841F} and \citet{2020MNRAS.493.1587H}, where the total signal $S_{i}$ is estimated as the sum of the signal of the sample galaxy ($S_{s}$) and the signal of the companions ($S_{c}$) weighted with two overlap factors:
\begin{eqnarray}
     S_{i} = S_{s} + \Sigma_{c}f_{1;c}f_{2;c}S_{c},
     \label{signal_with_confusion}
\end{eqnarray}
where the $f_{1}$ and $f_{2}$ model the overlap between the sample galaxy and its companion in angular and redshift space. The expected \hi\ mass of each companion is estimated using the $M_{\rm \hi}$ vs. galaxy optical diameter relation \citep{2011ApJ...732...93T}.
The true signal of the sample galaxy is obtained by:
\begin{eqnarray}
     S_{s} = S_{i} - \Sigma_{c}f_{1;c}f_{2:c}S_{c},
     \label{signal_without_confusion}
\end{eqnarray}
The confusion correction will be later applied to all confused galaxies in all stacking, by subtracting the signal from the companion before the final weighting is applied (Equation~\ref{weighted_spectrum}):
\begin{eqnarray}
     m^{s}_{\rm \hi}(\nu) = m^{i}_{\rm \hi}(\nu) - \Sigma_{c}m^{c}_{\rm \hi}(\nu),
     \label{mass_spectrum_confusion_c}
\end{eqnarray}
where $m^{s}_{\rm \hi}(\nu)$ refers to the true \hi mass spectrum from the sample galaxy, $m^{i}_{\rm \hi}(\nu)$ contains total \hi\ signal and $m^{c}_{\rm \hi}(\nu)$ is the estimated \hi\ signal from companion and $\int_{-\Delta\nu}^{\Delta\nu}\langle m_{\rm \hi}(\nu)\rangle d\nu = f_{1;c}f_{2:c}S_{c}$. The true \hi\ mass spectrum, $m^{s}_{\rm \hi}(\nu)$, is the spectrum we will use in the following stacking (in other words, $m_{\rm \hi,i}$ in Equation~\ref{weighted_spectrum}).

We do not see a clear dependence of the \hi-confused fraction on
redshift. Nearly the same stacking results and \hi\ properties-projected distance relations are obtained, when only galaxies classified as non confused are included in the analysis. However, to take full advantage of our large number statistics, in the rest of this paper we include confused galaxies and apply the method described above to correct the confusion.

\vspace*{-0.5cm}
\section{\hi\ content binned by group-centric radius}
\label{sec:results}
In this section, we present the \hi\ content vs. group-centric radius relations. In the Yang catalogue, the dark matter halos are defined as having an overdensity of 180, so the halo radius at which the average enclosed density is 180 times the critical density can be obtained from:
\begin{eqnarray}
      R_{180}=0.781h^{-1}\Mpc\left\lgroup\frac{M_{h}}{\Omega_{m}10^{14}h^{-1}\Msun}\right\rgroup^{1/3}(1+z_{\rm group})^{-1},
     \label{r180}
\end{eqnarray}
where $M_{h}$ is the mass of the group the satellite is located in and $z_{\rm group}$ is the redshift of the group centre. Using the technique described in Section~\ref{sec:script}, we can measure the \hi\ content in galaxies with different normalised projected group-centric radii.

\begin{figure}
    \centering
    \includegraphics[width=7.5cm]{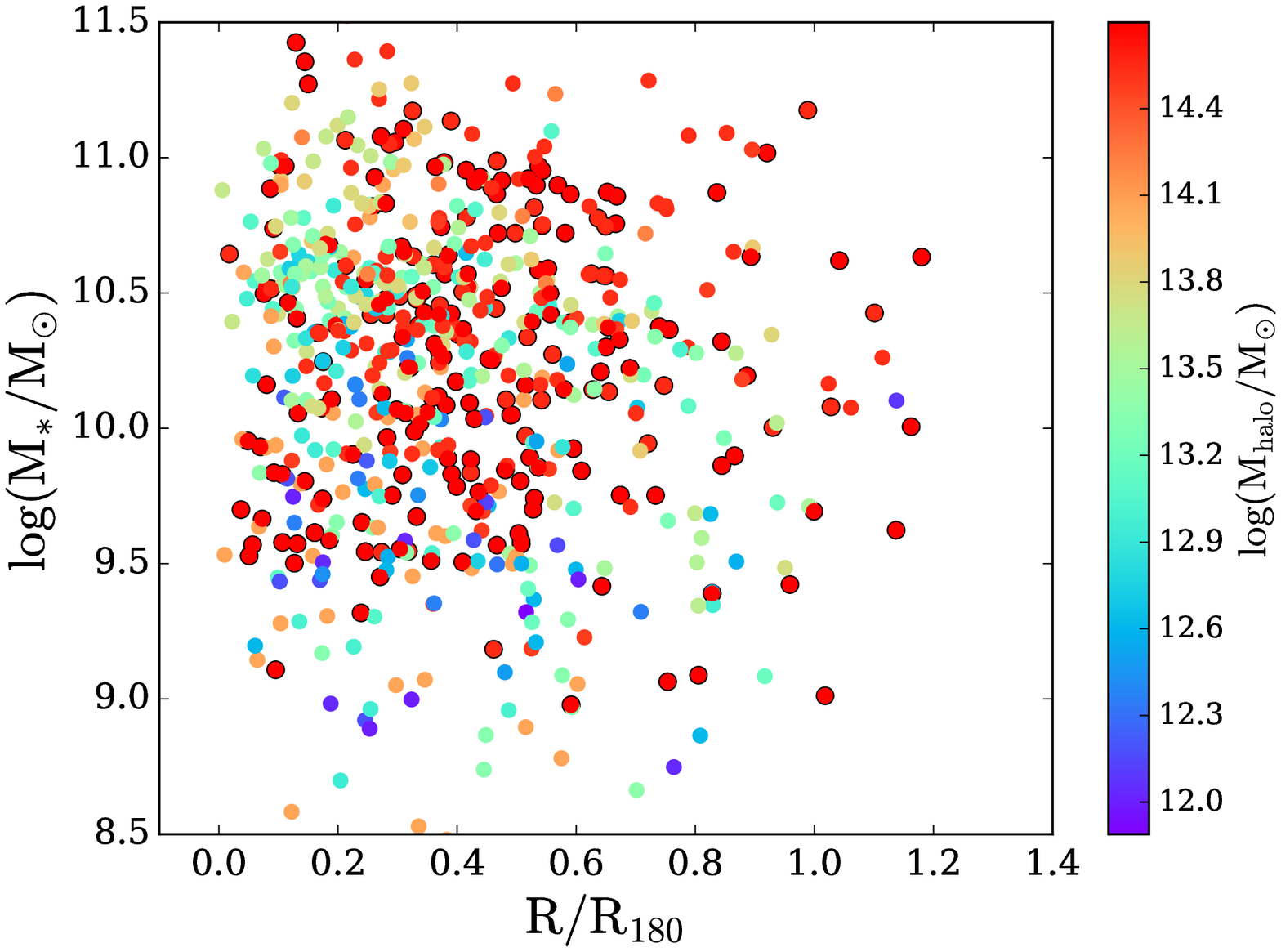}
    \includegraphics[width=7.5cm]{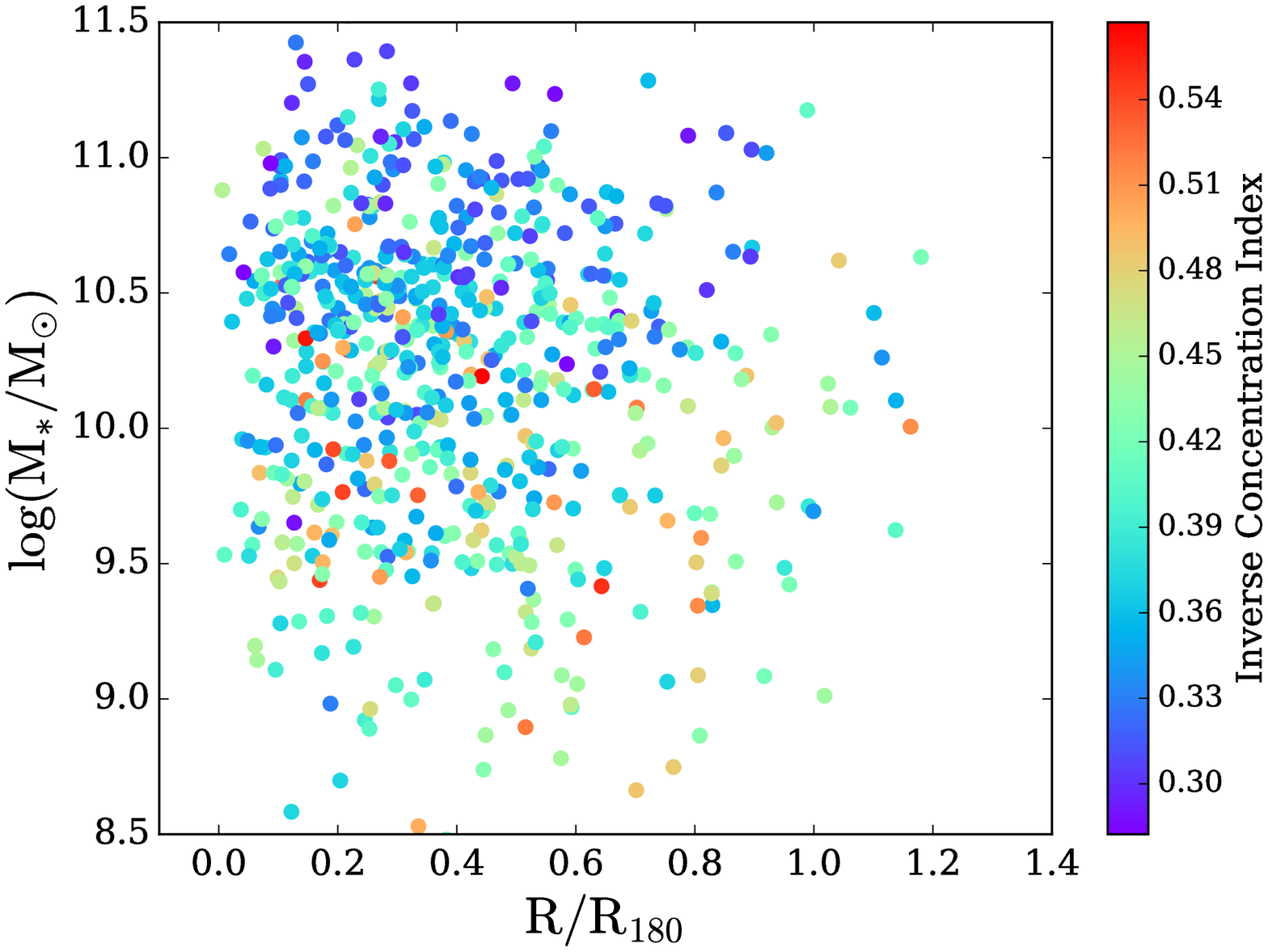}
    \caption{Stellar mass as a function of normalised projected group-centric radius for satellites, coloured according to the halo mass of the group the satellites are located in (upper panel) and their inverse concentration index (lower panel). The points with black circles refer to the satellites from the two largest groups.}
    \label{stellarmass_distance_diagram_satellite}
\end{figure}

Figure~\ref{stellarmass_distance_diagram_satellite} shows stellar mass as a function of normalised projected group-centric radius for satellites, coloured according to the halo mass of the group that the satellites are located in (upper panel) and their inverse concentration index (lower panel). Except for the satellites in the two largest groups (labeled with black circles), the high-mass galaxies are more likely to be found in high-mass groups. We should note that \hi\ mass, \hi\ gas fraction and \hi\ deficiency depend strongly on stellar mass, as shown in \citet{2016ApJ...824..110O} and \citet{2018MNRAS.476..875C}. If stellar mass is not controlled, the measured \hi\ properties vs. group-centric radius relations will affected by the relations between stellar mass and radius. In order to incorporate stellar mass as an additional independent variable and make use of as many galaxies as possible, we select the satellites by their stellar masses in the range $10^{10.0} < M_{\ast}/$M$_{\odot}  \leqslant 10^{11.5}$. The lower panel of Figure~\ref{stellarmass_distance_diagram_satellite} shows that the high-mass galaxies are more likely to have lower inverse concentration index and no clear relation between group-centric radius and morphology is found. In order to quantify the influence from galaxy type, we select the late-type satellites with inverse concentration index larger than 0.33 and carry out the measurement with only the late-type galaxies.

\begin{figure*}
\begin{multicols}{3}
      \includegraphics[width=6.cm,height=4.9cm]{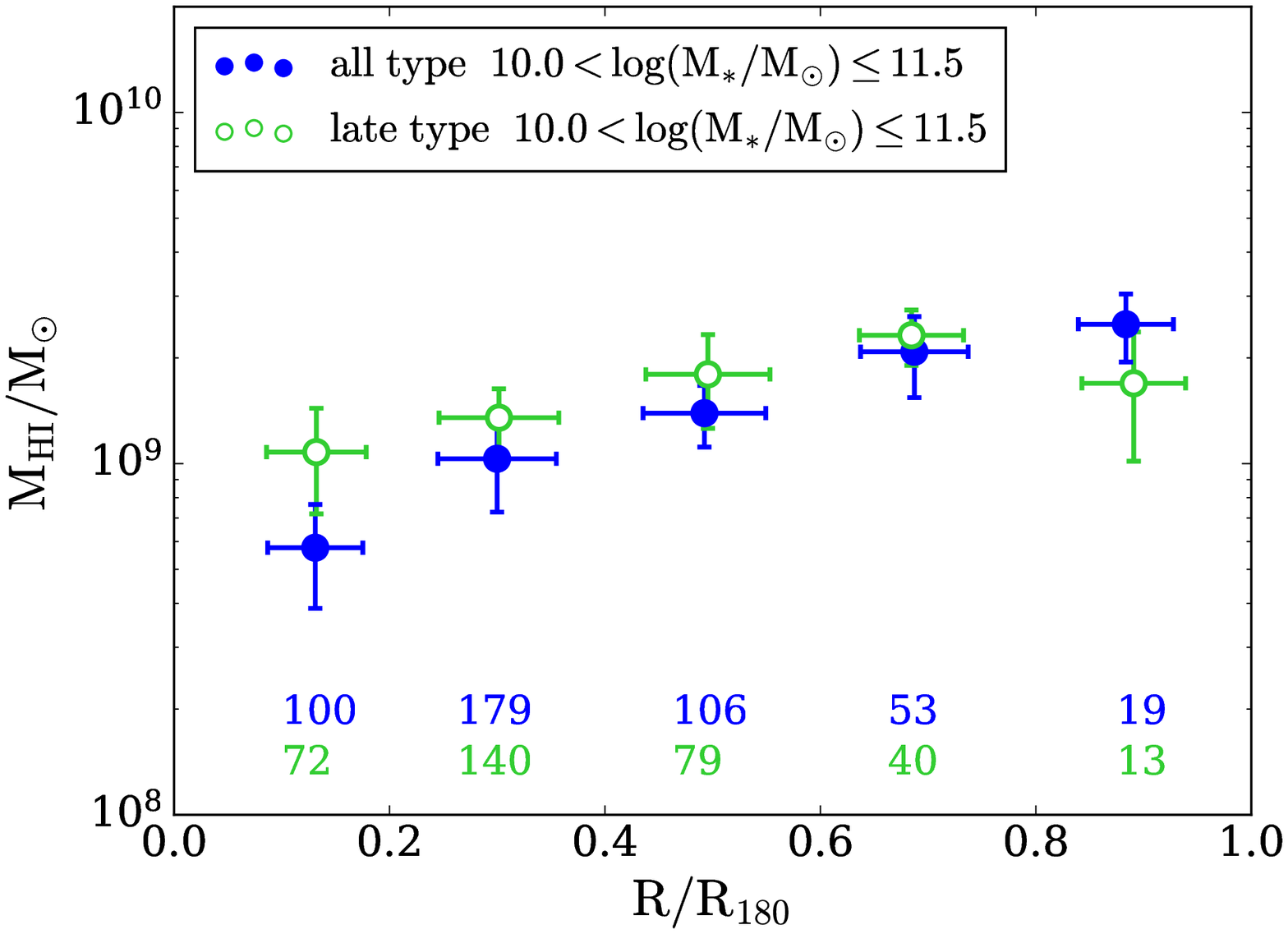}\par
      \hspace*{-0.3cm}
      \includegraphics[width=6.cm,height=4.9cm]{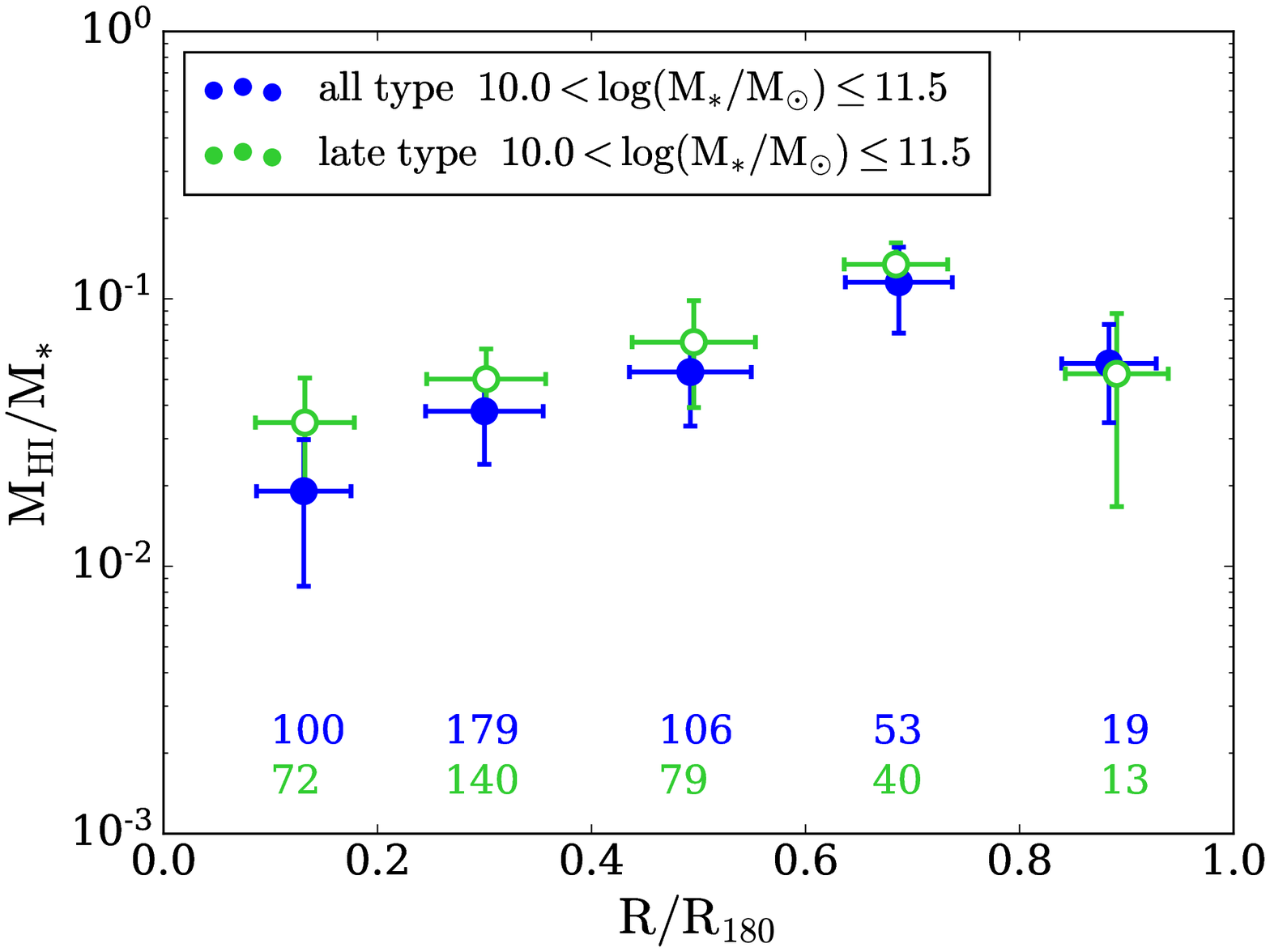}\par
      \hspace*{-0.5cm}
      \includegraphics[width=6.cm,height=4.9cm]{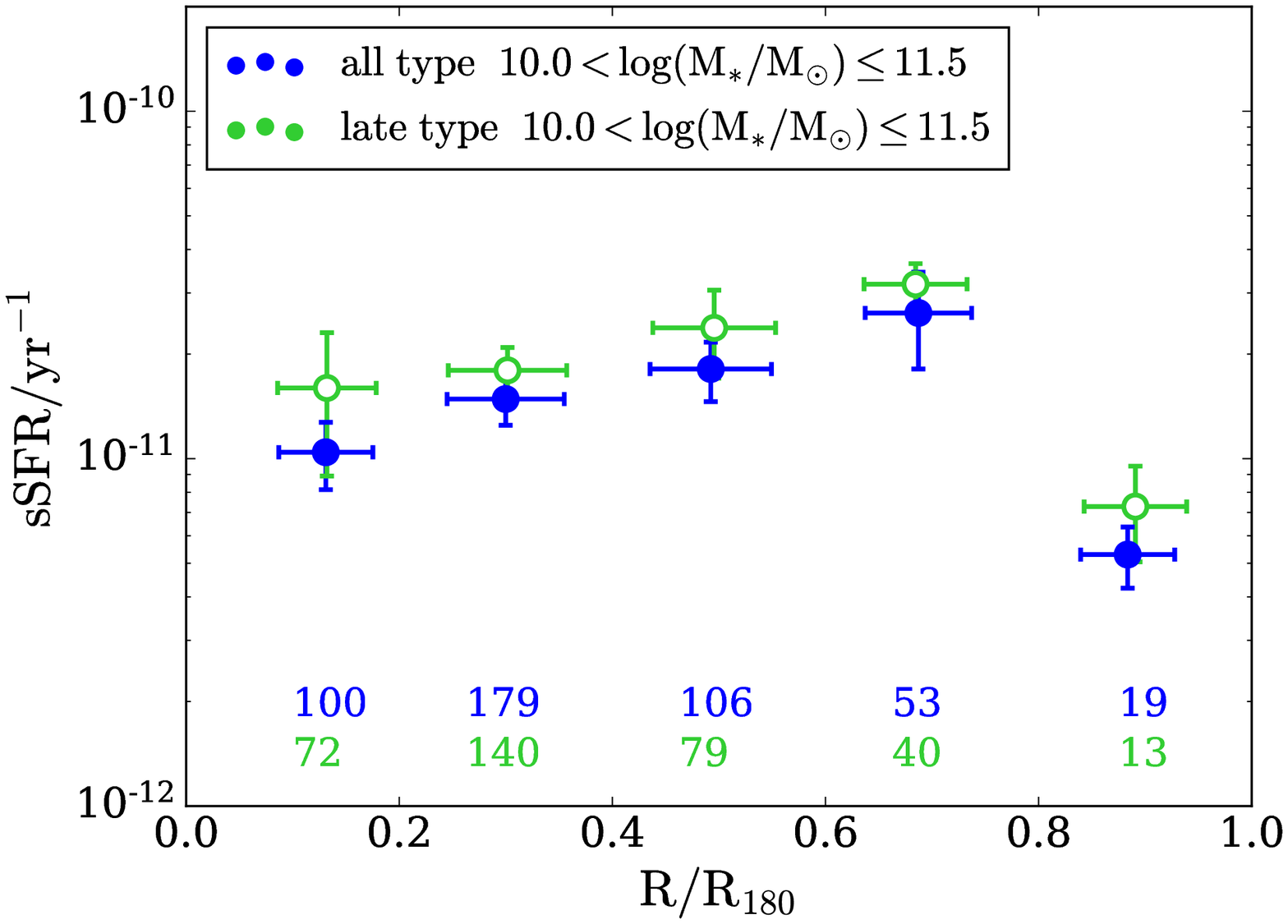}\par
\end{multicols}
\caption{Averaged \hi\ mass (left panel), \hi\ mass fraction (middle panel) and specific star formation rate (right panel) for the all-type (blue filled points) and the late-type (green open points) satellites with stellar mass: $10^{10.0} < M_{\ast} \leqslant 10^{11.5}$M$_{\odot}$, as a function of normalised projected group-centric radius. The error bars are estimated using jack-knife re-sampling. The numbers below the points show the number of galaxies in each radial bin. The corresponding values are presented in Table~\ref{distance_HIcontent_satellite_tab}.}
\label{distance_HIcontent_satellite}
\end{figure*}

The measured relations between \hi\ content and group-centric radius are shown in Figure~\ref{distance_HIcontent_satellite} and Table~\ref{distance_HIcontent_satellite_tab}. The stacked mass spectra for satellite galaxies are shown in Appendix ( Figure~\ref{stacked_spectra_satellite}). In  Table~\ref{distance_HIcontent_satellite_tab} we also present the inverse concentration index (labeled as $r_{50}/r_{90}$) and stacked stellar mass as a function of radius. The average stellar mass of satellites slightly decreases with increasing group distance for the first four bins. Given that \hi\ mass generally increases with stellar mass, if stellar mass is driving our results we should find higher gas masses closer to the group center. Conversely, for the stacks within $R/R_{180} \sim 0.8$, we find that at fixed stellar mass, satellite galaxies in the inner region lack \hi\ compared to galaxies in the outer region of each group. For the average \hi\ mass fraction, satellite galaxies in the inner region have lower values relative to satellites in the outer regions. The dependence of \hi\ content and \hi\ gas fraction on distance from the group centre for satellites at fixed stellar mass likely reflects the mechanism for removing gas in the inner regions.

For comparison, we stack the sSFR for the same sample using the same method, with \hi\ mass spectra replaced by sSFR, and show the results in the right panel of Figure~\ref{distance_HIcontent_satellite}. This shows that satellites in the outer regions of groups have higher sSFRs than those at small group-centric radii. The sSFR in the last bin is lowest, because those galaxies have highest averaged stellar mass.

For the late-type subsample, the same trends are found but with higher average \hi\ mass, \hi\ mass fraction and sSFR values in all except the last radial bins. We note that the average stellar mass of the late-type satellites is lower ($\sim 3.7 \times 10^{10}$M$_{\odot}$) and changes little with increasing group distance for the first four bins. For both of the all-type and the late-type sample, the average inverse concentration index of satellites changes little with increasing group distance, indicating that our results are not influenced by the morphology of galaxies.

\begin{table*}
 \caption{Stacked \hi\ properties as a function of group-centric radius for satellites with $10^{10.0} < M_{\ast} \leqslant 10^{11.5}$ M$_{\odot}$. We illustrate the results in Figure~\ref{distance_HIcontent_satellite}.}
 \label{distance_HIcontent_satellite_tab}
  \begin{tabular}{lccccccc}
   \hline
   Radius bin & Radius & Number of galaxies & $r_{50}/r_{90}$ & $\langle M_{\ast}\rangle$ & $\langle M_{\rm \hi}\rangle$ & $\langle M_{\rm \hi}/M_{\ast}\rangle$ & $\langle {\rm sSFR\rangle}$\\
    ($R_{180}$) & ($R_{180}$) & & & $(10^{10} h_{70}^{-2}$ M$_{\odot})$ & $(10^{9} h_{70}^{-2}$ M$_{\odot})$ & & (10$^{-11}$ yr$^{-1}$)\\
    \hline
    all type \\
   (0.0,0.2] & 0.13 $\pm$ 0.04 & 100 & 0.37 $\pm$ 0.05 &  4.8 $\pm$ 0.6 & 0.6 $\pm$ 0.2 & 0.02 $\pm$ 0.01 & 1.0 $\pm$ 0.2\\
   (0.2,0.4] & 0.30 $\pm$ 0.05 & 179 & 0.38 $\pm$ 0.06 & 4.5 $\pm$ 0.3 & 1.0 $\pm$ 0.3 & 0.04 $\pm$ 0.01 & 1.5 $\pm$ 0.2\\
   (0.4,0.6] & 0.49 $\pm$ 0.06 & 106 & 0.37 $\pm$ 0.06 & 4.3 $\pm$ 0.4 & 1.4 $\pm$ 0.3 & 0.05 $\pm$ 0.02 & 1.8 $\pm$ 0.3\\
   (0.6,0.8] & 0.69 $\pm$ 0.05 & 53 & 0.38 $\pm$ 0.06 & 3.6 $\pm$ 0.5 & 2.1 $\pm$ 0.5 & 0.12 $\pm$ 0.04 & 2.6 $\pm$ 0.8\\
   (0.8,1.0] & 0.88 $\pm$ 0.04 & 19 & 0.37 $\pm$ 0.06 & 5.3 $\pm$ 1.1 & 2.5 $\pm$ 0.5 & 0.06 $\pm$ 0.02 & 0.5 $\pm$ 0.1\\
   \hline
   late type \\
   (0.0,0.2] & 0.13 $\pm$ 0.05 & 72 & 0.39 $\pm$ 0.05 & 3.6 $\pm$ 0.3 & 1.1 $\pm$ 0.3 & 0.03 $\pm$ 0.02 & 1.6 $\pm$ 0.5\\
   (0.2,0.4] & 0.30 $\pm$ 0.05 & 140 & 0.39 $\pm$ 0.05 & 3.8 $\pm$ 0.3 & 1.4 $\pm$ 0.3 & 0.05 $\pm$ 0.01 & 1.8 $\pm$ 0.3\\
   (0.4,0.6] & 0.49 $\pm$ 0.06 & 79 & 0.40 $\pm$ 0.05 & 3.7 $\pm$ 0.4 & 1.8 $\pm$ 0.5 & 0.07 $\pm$ 0.03 & 2.4 $\pm$ 0.6\\
   (0.6,0.8] & 0.68 $\pm$ 0.05 & 40 & 0.40 $\pm$ 0.05 & 3.3 $\pm$ 0.4 & 2.3 $\pm$ 0.4 & 0.13 $\pm$ 0.03 & 3.2 $\pm$ 0.5\\
   (0.8,1.0] & 0.89 $\pm$ 0.05 & 13 & 0.40 $\pm$ 0.05 & 4.2 $\pm$ 1.1 & 1.7 $\pm$ 0.7 & 0.05 $\pm$ 0.03 & 0.7 $\pm$ 0.2\\
   \hline
  \end{tabular}
\end{table*}

\subsection{Environmental dependence}
\label{sec:Environmental dependence}
To study the physical mechanism behind the \hi\ properties vs. group-centric radius relations, we further bin the satellites by the halo mass of groups in which they are located, or by projected density, and compare the results obtained with these two different environmental metrics.

\subsubsection{Group Halo mass}
We split the satellites in the stellar mass range $10^{10.0} < M_{\ast} \leqslant 10^{11.5}$M$_{\odot}$ into two sub-samples according to the mass of the group in which they reside: $ M_{\rm halo} \leqslant 10^{13.5}h^{-1}$M$_{\odot}$ and $M_{\rm halo} > 10^{13.5}h^{-1}$M$_{\odot}$. There are 90 (79 for the late-type subsample) groups and 25 (24 for the late-type subsample) groups in these two sub-samples, respectively. The low-mass and high-mass group sub-samples have an averaged halo mass of $ (1.41 \pm 1.07) \times10^{13} M_{\odot}$ and $(1.54 \pm 1.44) \times10^{14} M_{\odot}$, respectively ($(1.40 \pm 1.07) \times10^{13} M_{\odot}$ and $(1.58 \pm 1.45) \times10^{14} M_{\odot}$ for the late-type subsample).

\begin{figure*}
\begin{multicols}{3}
\includegraphics[width=6.cm,height=4.9cm]{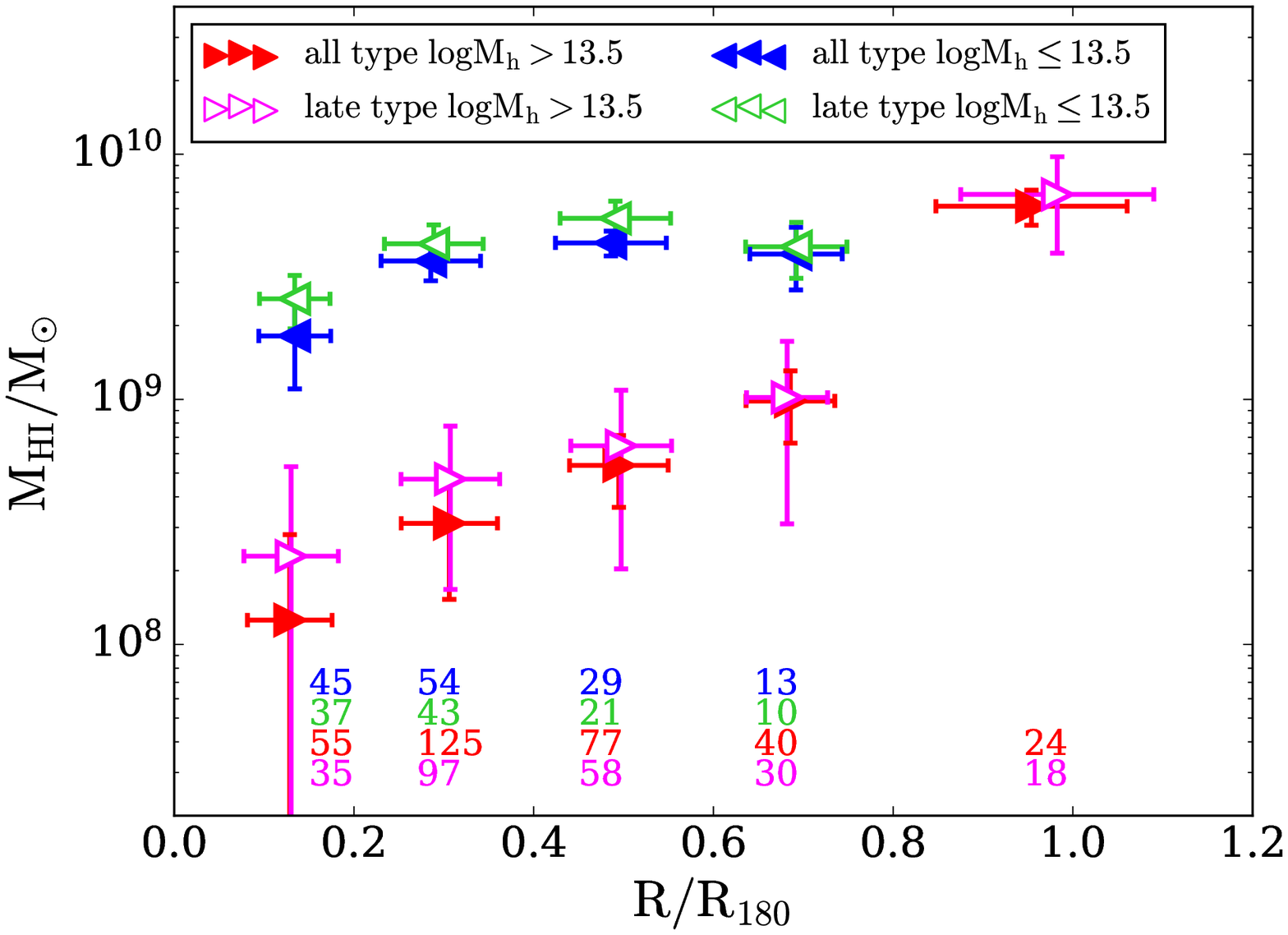}\par
\hspace*{-0.3cm}
\includegraphics[width=6.cm,height=4.9cm]{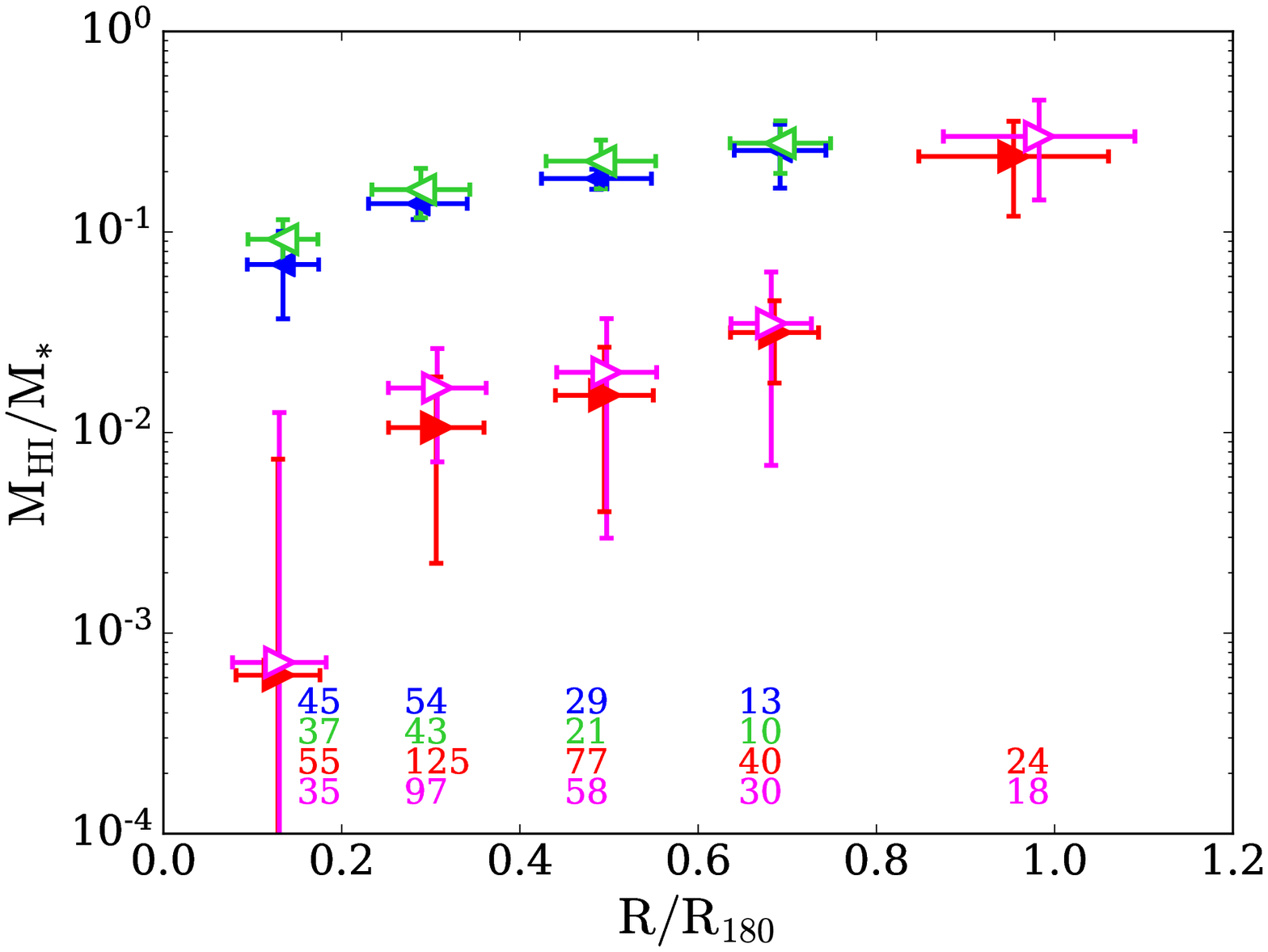}\par
\hspace*{-0.3cm}
\includegraphics[width=6.cm,height=4.9cm]{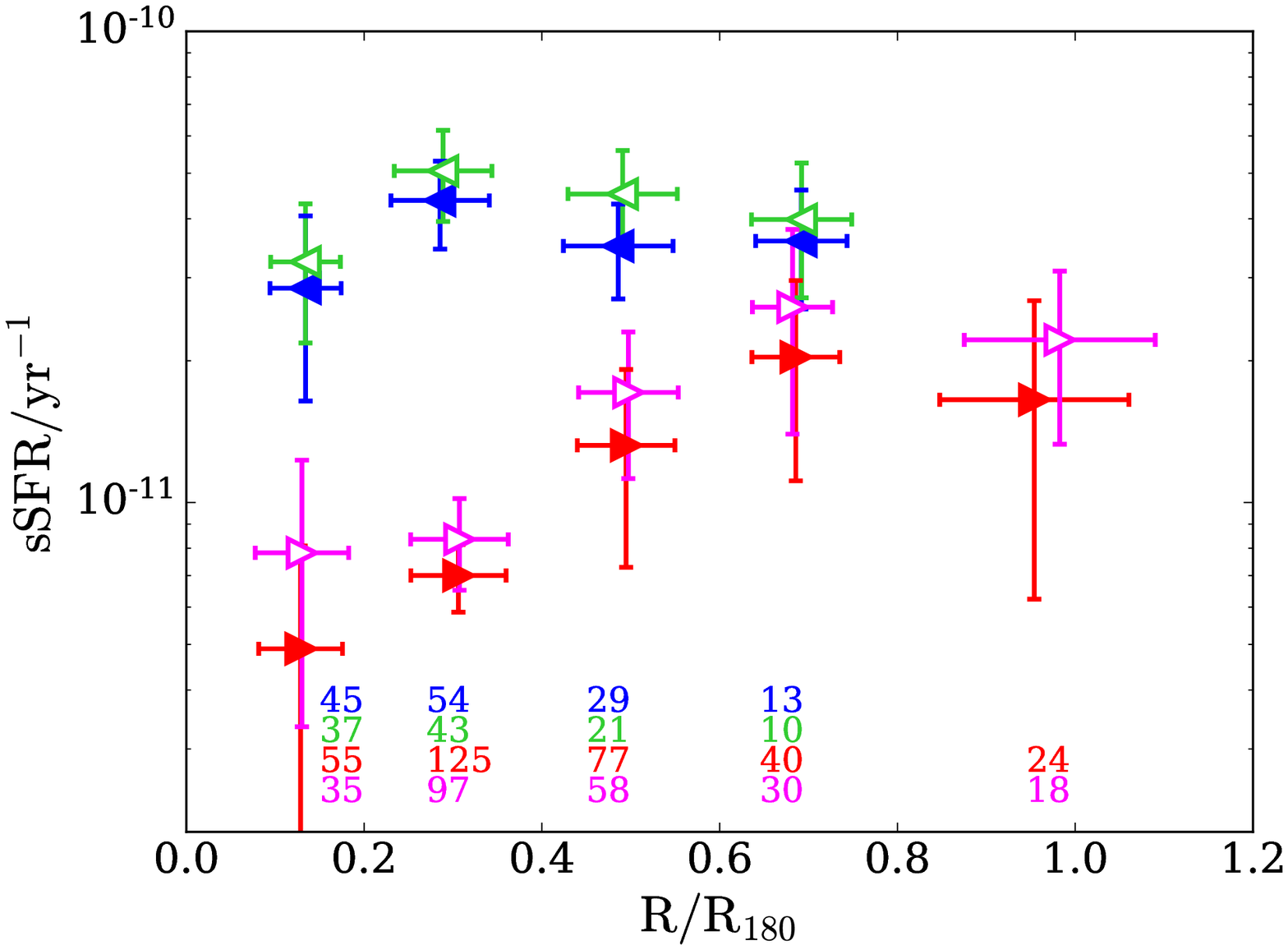}\par
\end{multicols}
\caption{Same as Figure~\ref{distance_HIcontent_satellite}, with satellite galaxies in $10^{10.0} < M_{\ast} \leqslant 10^{11.5}$M$_{\odot}$ divided into two group halo mass bins, above and below $10^{13.5}h^{-1}$M$_{\odot}$ (right-pointing and left-pointing triangles), respectively. The results from the all-type and the late-type satellites are labeled as filled and open points, respectively. The corresponding values are presented in Table~\ref{distance_HIcontent_satellite_group_tab}.}
\label{distance_HIcontent_satellite_group}
\end{figure*}

We show the stacked \hi\ mass, \hi\ gas fraction and sSFR in different radial bins in Figure~\ref{distance_HIcontent_satellite_group} and Table~\ref{distance_HIcontent_satellite_group_tab}. The average stellar mass from stacks of sub-samples in the first several radial bins are similar, both for high-mass groups and low-mass groups. For the high-mass group sub-sample, the last radial bin was extended to 1.2 $R_{180}$ to contain more galaxies and improve statistics.

We find a monotonic decrease of \hi\ mass in satellite galaxies with decreasing distance from the centre of the group for both low- and high-mass groups, with high-mass groups showing a more dramatic drop in gas content.  
 
For satellites in high-mass groups, the \hi\ content residing in the most inner parts ($R \sim 0.13\ R_{180}$) of groups is $\sim$ 60 times smaller than that in the outer parts ($R \sim 0.95\ R_{180}$). It is worth stressing that, in the inner parts of high-mass groups, the \hi\ content residing in satellites is less than in low-mass groups, even though the average stellar masses are higher in the former. This indicates that the gas removal process is more active in higher-mass groups.

Similar trends apply to \hi\ gas fraction and sSFR, although in the case of gas fraction the trend is noisier due to the fact that for two more radial bins the stacking results are more uncertain. Specifically, we do not detect \hi\ in the first radial bin, and we have marginal detections (signal-to-noise lower than 3) in the third and fourth radial bins. 

Similar trends are found for the late-type subsample at fixed stellar mass, but with higher average \hi\ mass, \hi\ mass fraction and sSFR values in nearly all radial bins.

\begin{table*}
 \caption{Stacked \hi\ properties as a function of group-centric radius for satellites with 10$^{10.0}$M$_{\odot}$ < $M_{\ast}$ $\leqslant$ 10$^{11.5}$M$_{\odot}$ in two bins of group halo mass. We illustrate the results in Figure~\ref{distance_HIcontent_satellite_group}.}
 \label{distance_HIcontent_satellite_group_tab}
  \begin{tabular}{lccccccc}
   \hline
   Radial bins & Radius & Number of galaxies & $r_{50}/r_{90}$ & $\langle M_{\ast}\rangle$ & $\langle M_{\rm \hi}\rangle$ & $\langle M_{\rm \hi}/M_{\ast}\rangle$ & $\langle sSFR\rangle$\\
    ($R_{180}$) & ($R_{180}$) & & & $(10^{10} h_{70}^{-2}$ M$_{\odot})$ & $(10^{9} h_{70}^{-2}$ M$_{\odot})$ & & (10$^{-11}$/yr$^{-1}$)\\
    \hline
    \multicolumn{8}{l}{low-mass group: $M_{\rm halo} \leqslant 10^{13.5}h^{-1}$M$_{\odot}$, all type}\\
   (0.0,0.2] & 0.13 $\pm$ 0.04 & 45 & 0.39 $\pm$ 0.06 & 3.4 $\pm$ 0.2 & 1.8 $\pm$ 0.7 & 0.07 $\pm$ 0.03 & 2.8 $\pm$ 1.2\\
   (0.2,0.4] & 0.29 $\pm$ 0.05 & 54 & 0.39 $\pm$ 0.06 & 3.6 $\pm$ 0.7 & 3.7 $\pm$ 0.6 & 0.14 $\pm$ 0.02 & 4.4 $\pm$ 0.9\\
   (0.4,0.6] & 0.49 $\pm$ 0.06 & 29 & 0.38 $\pm$ 0.06 & 3.4 $\pm$ 0.7 & 4.3 $\pm$ 0.5 & 0.18 $\pm$ 0.02 & 3.5 $\pm$ 0.8\\
   (0.6,0.8] & 0.69 $\pm$ 0.05 & 13 & 0.40 $\pm$ 0.07 & 2.2 $\pm$ 0.2 & 3.9 $\pm$ 1.1 & 0.25 $\pm$ 0.08 & 3.6 $\pm$ 1.0\\
   \hline
    \multicolumn{8}{l}{high-mass group: $M_{\rm halo} > 10^{13.5}h^{-1}$M$_{\odot}$, all type}\\
   (0.0,0.2] & 0.13 $\pm$ 0.05 & 55 & 0.35 $\pm$ 0.04 & 5.2 $\pm$ 0.8 & 0.1 $\pm$ 0.2 & 0.001 $\pm$ 0.007 & 0.5 $\pm$ 0.3\\
   (0.2,0.4] & 0.31 $\pm$ 0.05 & 125 & 0.37 $\pm$ 0.05 & 5.0 $\pm$ 0.3 & 0.3 $\pm$ 0.2 & 0.011 $\pm$ 0.008 & 0.7 $\pm$ 0.1\\
   (0.4,0.6] & 0.49 $\pm$ 0.05 & 77 & 0.37 $\pm$ 0.05 & 4.7 $\pm$ 0.3 & 0.5 $\pm$ 0.2 & 0.015 $\pm$ 0.011 & 1.3 $\pm$ 0.6\\
   (0.6,0.8] & 0.69 $\pm$ 0.05 & 40 & 0.37 $\pm$ 0.05 & 4.1 $\pm$ 0.6 & 1.0 $\pm$ 0.3 & 0.032 $\pm$ 0.014 & 2.0 $\pm$ 0.9\\
   (0.8,1.2] & 0.95 $\pm$ 0.11 & 24 & 0.38 $\pm$ 0.06 & 4.8 $\pm$ 0.3 & 6.1 $\pm$ 1.0 & 0.238 $\pm$ 0.118 & 1.7 $\pm$ 1.0\\
   \hline
    \multicolumn{8}{l}{low-mass group: $M_{\rm halo} \leqslant 10^{13.5}h^{-1}$M$_{\odot}$, late type}\\
   (0.0,0.2] & 0.13 $\pm$ 0.04 & 37 & 0.41 $\pm$ 0.05 & 3.2 $\pm$ 0.3 & 2.6 $\pm$ 0.6 & 0.09 $\pm$ 0.02 & 3.2 $\pm$ 1.1\\
   (0.2,0.4] & 0.29 $\pm$ 0.05 & 43 & 0.41 $\pm$ 0.06 & 3.5 $\pm$ 0.3 & 4.3 $\pm$ 0.8 & 0.16 $\pm$ 0.04 & 5.1 $\pm$ 1.1\\
   (0.4,0.6] & 0.49 $\pm$ 0.06 & 21 & 0.41 $\pm$ 0.05 & 3.0 $\pm$ 0.4 & 5.5 $\pm$ 0.9 & 0.23 $\pm$ 0.06 & 4.5 $\pm$ 1.1\\
   (0.6,0.8] & 0.69 $\pm$ 0.05 & 10 & 0.43 $\pm$ 0.06 & 2.2 $\pm$ 0.3 & 4.2 $\pm$ 1.0 & 0.28 $\pm$ 0.08 & 4.0 $\pm$ 1.2\\
   \hline
    \multicolumn{8}{l}{high-mass group: $M_{\rm halo} > 10^{13.5}h^{-1}$M$_{\odot}$, late type}\\
   (0.0,0.2] & 0.13 $\pm$ 0.05 & 35 & 0.37 $\pm$ 0.04 & 3.7 $\pm$ 0.6 & 0.2 $\pm$ 0.3 & 0.001 $\pm$ 0.012 & 0.8 $\pm$ 0.4\\
   (0.2,0.4] & 0.31 $\pm$ 0.05 & 97 & 0.39 $\pm$ 0.04 & 4.1 $\pm$ 0.2 & 0.5 $\pm$ 0.3 & 0.017 $\pm$ 0.009 & 0.8 $\pm$ 0.2\\
   (0.4,0.6] & 0.50 $\pm$ 0.06 & 58 & 0.39 $\pm$ 0.04 & 4.0 $\pm$ 0.4 & 0.6 $\pm$ 0.4 & 0.020 $\pm$ 0.017 & 1.7 $\pm$ 0.6\\
   (0.6,0.8] & 0.68 $\pm$ 0.04 & 30 & 0.39 $\pm$ 0.04 & 3.6 $\pm$ 0.9 & 1.0 $\pm$ 0.7 & 0.035 $\pm$ 0.028 & 2.6 $\pm$ 1.2\\
   (0.8,1.2] & 0.98 $\pm$ 0.11 & 18 & 0.40 $\pm$ 0.05 & 3.8 $\pm$ 1.3 & 6.8 $\pm$ 3.0 & 0.299 $\pm$ 0.155 & 2.2 $\pm$ 0.9\\
   \hline
  \end{tabular}
\end{table*}

\subsubsection{Projected Density}
We also consider the local density as an environment metric and measure the \hi\ content vs. group-centric radius relations in sub-samples of different local projected densities. To calculate the local projected density ($\Sigma$), we count the number of neighbouring SDSS galaxies per projected Mpc$^{2}$. The galaxies are limited to those within the redshift range of each group. 
We then split the satellites into two sub-samples with local density $\Sigma > 5$ and $\Sigma \leq 5$ Mpc$^{-2}$. For stacking purposes, we again only use the satellites with $10^{10.0} < M_{\ast} \leqslant 10^{11.5}$M$_{\odot}$ and limit our measurement to the late-type galaxies. The low-density and high-density sub-samples have an averaged local density of $(2.87 \pm 0.85)$ Mpc$^{-2}$ and $(13.07 \pm 9.52)$ Mpc$^{-2}$ ($(2.88 \pm 0.81)$ Mpc$^{-2}$ and $(12.33 \pm 8.83)$ Mpc$^{-2}$ for the late-type satellites), respectively.

\begin{figure*}
\begin{multicols}{3}
\includegraphics[width=6.cm,height=4.9cm]{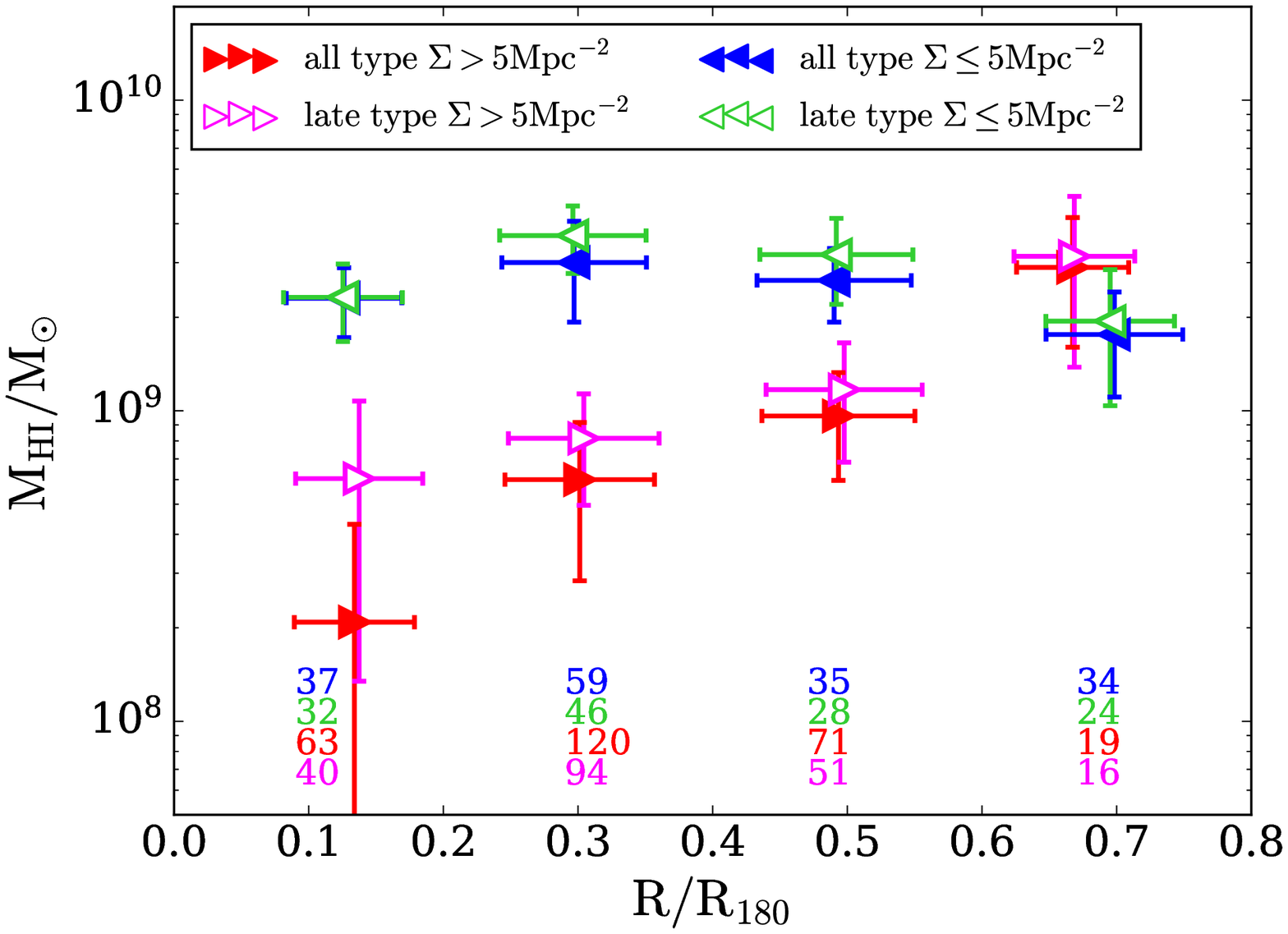}\par
\hspace*{-0.3cm}
\includegraphics[width=6.cm,height=4.9cm]{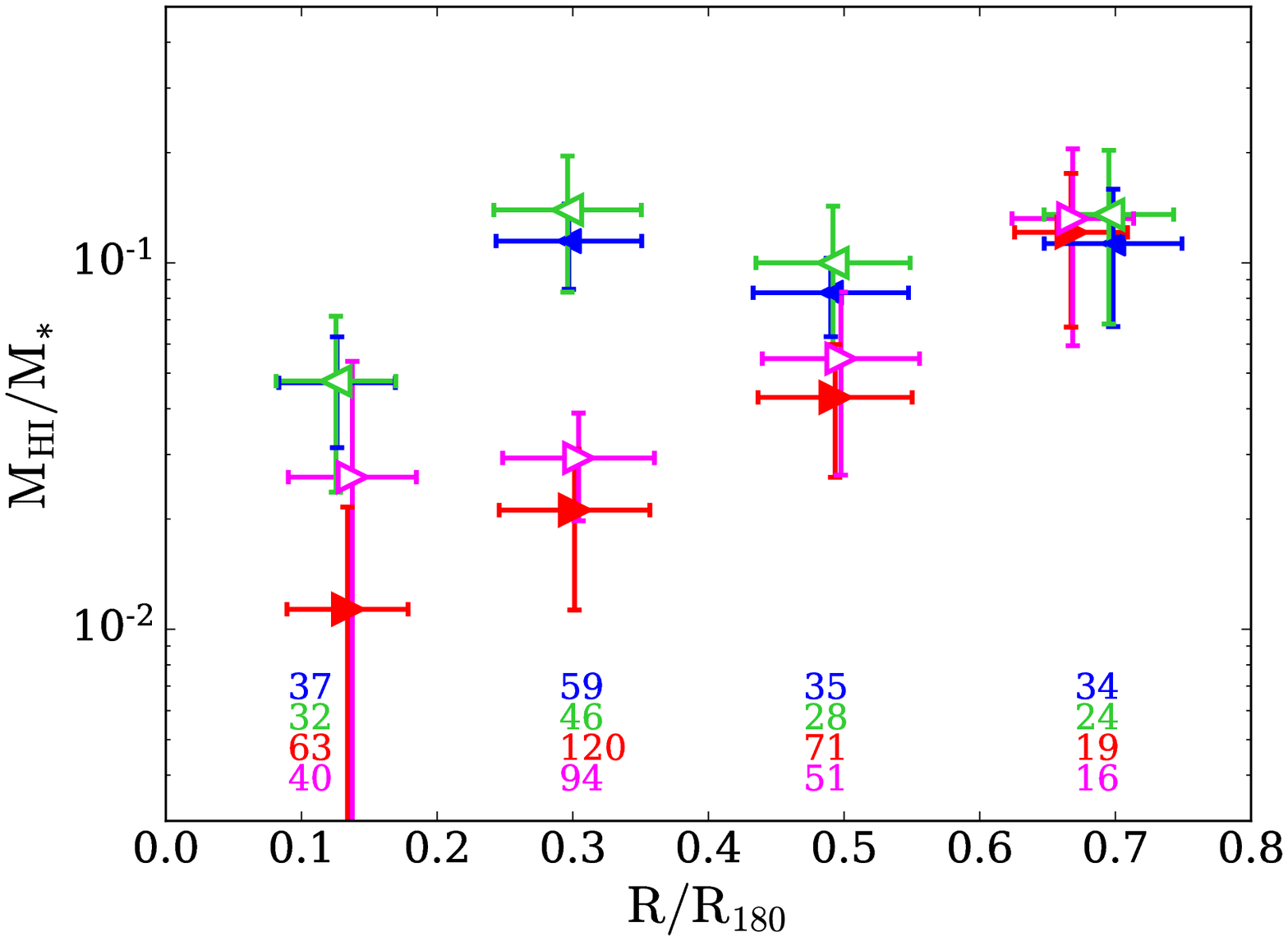}\par
\hspace*{-0.3cm}
\includegraphics[width=6.cm,height=4.9cm]{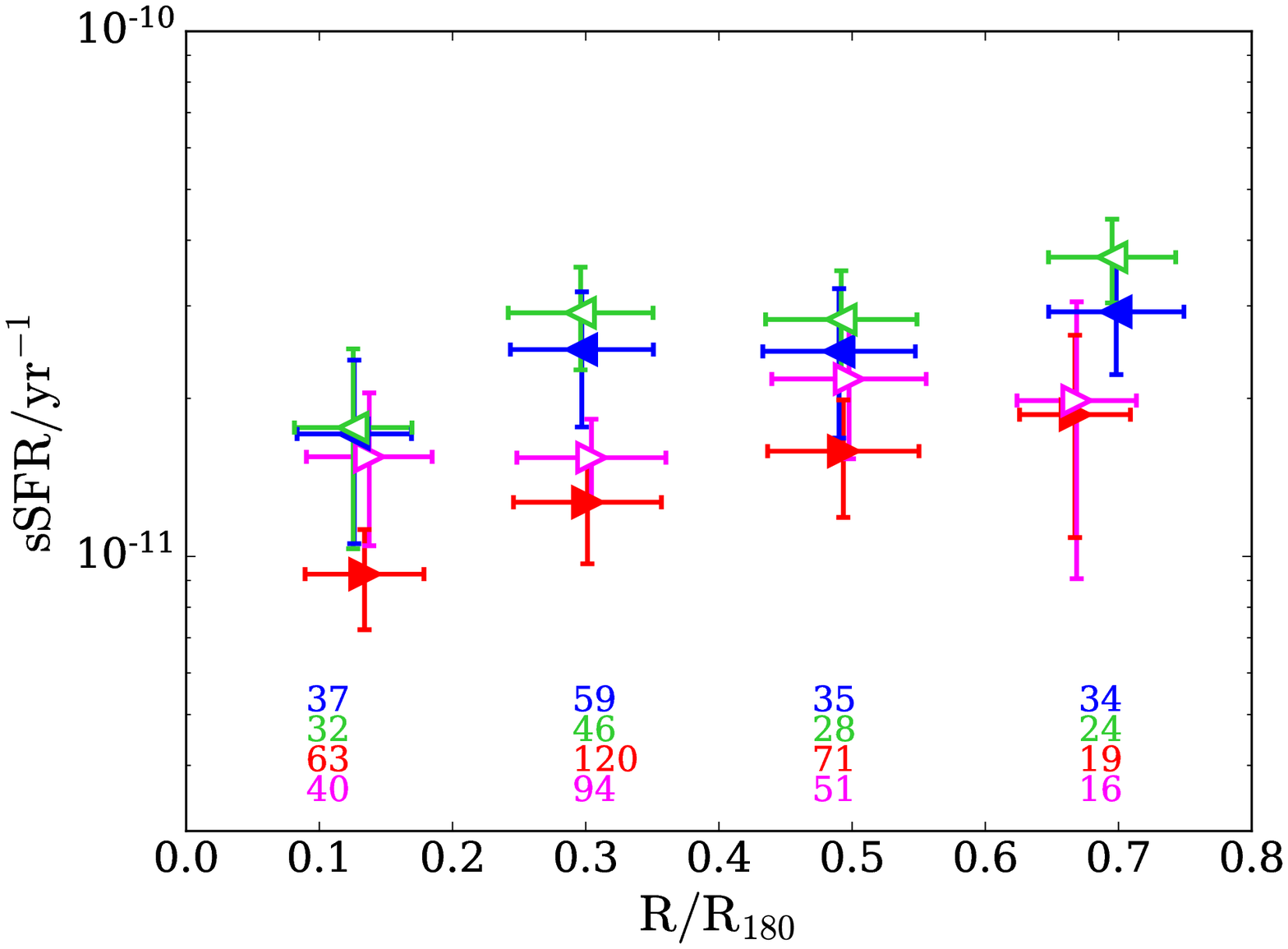}\par
\end{multicols}
\caption{Same as Figure~\ref{distance_HIcontent_satellite}, with satellite galaxies in $10^{10.0} < M_{\ast} \leqslant 10^{11.5}$M$_{\odot}$ divided into two local density bins, above and below 5 Mpc$^{-2}$ (right-pointing and left-pointing triangles), respectively. The corresponding values are presented in Table~\ref{distance_HIcontent_satellite_density_tab}.}
\label{distance_HIcontent_satellite_density}
\end{figure*}

We stack the \hi\ mass, \hi\ gas fraction and sSFR of satellites in different radial bins and show the results in Figure~\ref{distance_HIcontent_satellite_density} and Table~\ref{distance_HIcontent_satellite_density_tab}. Low densities ($\Sigma \leq$5 Mpc$^{-2}$) and high densities ($\Sigma $>5 Mpc$^{-2}$) are indicated in blue and red, respectively. Table~\ref{distance_HIcontent_satellite_density_tab} shows that the average stellar mass changes little with group-centric radius, for the first three radial bins.

The relation between average \hi\ mass and projected radius clearly depends on the local densities. At a fixed stellar mass, inner satellites in high-density regions lack \hi\ relative to galaxies in the outer region of each density group. However,  satellites in low-density regions show no change of \hi\ content with radius. Similar results are obtained if we plot gas fractions instead of gas masses. This suggests that \hi\ removal is radially dependent only in relatively high density regions within groups. 

Stacks of sSFR show that the satellites in higher-density regions have lower sSFR at fixed stellar mass. For satellites in lower density environments, a radial trend with group-centric radius may still be present, but only within $R$ $\sim$ 0.3 $R_{180}$. 

Similar results are obtained for the late-type subsample, but with more moderate increasing trends with radius.

\begin{table*}
 \caption{Stacked \hi\ properties as a function of group-centric radius for satellites with 10$^{10.0}$M$_{\odot}$ < $M_{\ast}$ $\leqslant$ 10$^{11.5}$M$_{\odot}$ in two bins of local density. We illustrate the results in Figure~\ref{distance_HIcontent_satellite_density}.}
 \label{distance_HIcontent_satellite_density_tab}
  \begin{tabular}{lccccccc}
   \hline
   Radial bins & Radius & Number of galaxies & $r_{50}/r_{90}$ & $\langle M_{\ast}\rangle$ & $\langle M_{\rm \hi}\rangle$ & $\langle M_{\rm \hi}/M_{\ast}\rangle$ & $\langle sSFR\rangle$\\
    ($R_{180}$) & ($R_{180}$) & & &  $(10^{10} h_{70}^{-2}$ M$_{\odot})$ & $(10^{9} h_{70}^{-2}$ M$_{\odot})$ & & (10$^{-11}$/yr$^{-1}$)\\
    \hline
    \multicolumn{8}{l}{low density: $\Sigma \leqslant$ 5 Mpc$^{-2}$, all type}\\
   (0.0,0.2] & 0.13 $\pm$ 0.04 & 37 & 0.39 $\pm$ 0.06 & 4.1 $\pm$ 0.5 & 2.3 $\pm$ 0.6 & 0.05 $\pm$ 0.02 & 1.7 $\pm$ 0.6\\
   (0.2,0.4] & 0.30 $\pm$ 0.05 & 59 & 0.39 $\pm$ 0.06 & 4.5 $\pm$ 0.3 & 3.0 $\pm$ 1.1 & 0.11 $\pm$ 0.03 & 2.5 $\pm$ 0.7\\
   (0.4,0.6] & 0.49 $\pm$ 0.06 & 35 & 0.38 $\pm$ 0.05 & 4.0 $\pm$ 0.6 & 2.6 $\pm$ 0.7 & 0.08 $\pm$ 0.02 & 2.5 $\pm$ 0.8\\
   (0.6,0.8] & 0.70 $\pm$ 0.05 & 34 & 0.38 $\pm$ 0.06 & 3.5 $\pm$ 0.5 & 1.8 $\pm$ 0.6 & 0.11 $\pm$ 0.04 & 2.9 $\pm$ 0.7\\
    \hline
    \multicolumn{8}{l}{high density: $\Sigma > $ 5 Mpc$^{-2}$, all type}\\
   (0.0,0.2] & 0.13 $\pm$ 0.04 & 63 & 0.36 $\pm$ 0.04 & 4.9 $\pm$ 0.8 & 0.2 $\pm$ 0.2 & 0.011 $\pm$ 0.010 & 0.9 $\pm$ 0.2\\
   (0.2,0.4] & 0.30 $\pm$ 0.05 & 120 & 0.37 $\pm$ 0.05 & 4.5 $\pm$ 0.4 & 0.6 $\pm$ 0.3 & 0.021 $\pm$ 0.009 & 1.3 $\pm$ 0.3\\
   (0.4,0.6] & 0.49 $\pm$ 0.06 & 71 & 0.37 $\pm$ 0.06 & 4.5 $\pm$ 0.3 & 1.0 $\pm$ 0.4 & 0.043 $\pm$ 0.017 & 1.6 $\pm$ 0.4\\
   (0.6,0.8] & 0.67 $\pm$ 0.04 & 19 & 0.38 $\pm$ 0.05 & 3.3 $\pm$ 0.1 & 2.9 $\pm$ 1.2 & 0.121 $\pm$ 0.054 & 1.9 $\pm$ 0.7\\
   \hline
    \multicolumn{8}{l}{low density: $\Sigma \leqslant$ 5 Mpc$^{-2}$, late type}\\
   (0.0,0.2] & 0.13 $\pm$ 0.04 & 32 & 0.41 $\pm$ 0.05 & 4.2 $\pm$ 0.6 & 2.3 $\pm$ 0.6 & 0.05 $\pm$ 0.02 & 1.8 $\pm$ 0.7\\
   (0.2,0.4] & 0.30 $\pm$ 0.05 & 46 & 0.40 $\pm$ 0.05 & 4.1 $\pm$ 0.7 & 3.7 $\pm$ 0.9 & 0.14 $\pm$ 0.06 & 2.9 $\pm$ 0.6\\
   (0.4,0.6] & 0.49 $\pm$ 0.06 & 28 & 0.40 $\pm$ 0.04 & 3.2 $\pm$ 0.3 & 3.2 $\pm$ 0.9 & 0.10 $\pm$ 0.04 & 2.8 $\pm$ 0.7\\
   (0.6,0.8] & 0.70 $\pm$ 0.05 & 24 & 0.40 $\pm$ 0.05 & 3.1 $\pm$ 0.6 & 1.9 $\pm$ 0.9 & 0.14 $\pm$ 0.06 & 3.7 $\pm$ 0.7\\
   \hline
    \multicolumn{8}{l}{high density: $\Sigma > $ 5 Mpc$^{-2}$, late type}\\
   (0.0,0.2] & 0.14 $\pm$ 0.05 & 40 & 0.38 $\pm$ 0.04 & 3.2 $\pm$ 0.4 & 0.6 $\pm$ 0.5 & 0.026 $\pm$ 0.028 & 1.5 $\pm$ 0.5\\
   (0.2,0.4] & 0.30 $\pm$ 0.05 & 94 & 0.39 $\pm$ 0.05 & 3.6 $\pm$ 0.3 & 0.8 $\pm$ 0.3 & 0.029 $\pm$ 0.010 & 1.5 $\pm$ 0.3\\
   (0.4,0.6] & 0.50 $\pm$ 0.06 & 51 & 0.39 $\pm$ 0.05 & 4.0 $\pm$ 0.4 & 1.2 $\pm$ 0.5 & 0.055 $\pm$ 0.028 & 2.2 $\pm$ 0.6\\
   (0.6,0.8] & 0.67 $\pm$ 0.04 & 16 & 0.39 $\pm$ 0.04 & 3.5 $\pm$ 0.2 & 3.1 $\pm$ 1.7 & 0.132 $\pm$ 0.072 & 2.0 $\pm$ 1.0\\
   \hline

  \end{tabular}
\end{table*}

\subsection{Extending to Larger Distances}
\label{sec:larger_distance}
Using stacking and multiple linear regressions, \citet{2016ApJ...824..110O} presented the \hi\ content distribution in nearby groups and clusters measured by the 70$\%$-complete ALFALFA survey. They found that at fixed stellar mass, the late-type galaxies in the inner parts of groups lack \hi\ compared with galaxies in a control region extending to 4.0 Mpc surrounding each group. The effect is detectable in groups with $M_{200} \lesssim 10^{14.5}$M$_{\odot}$ as well as in clusters ($M_{200} \gtrsim 10^{14.5}$M$_{\odot}$), which indicates pre-processing of \hi\ gas in intermediate-density isolated groups.

Motivated by these findings, we extend our analysis to higher group-centric radii by searching for isolated galaxies around each group. 
The distribution of galaxies in groups depends on their orbital and infall history, and can be modelled by phase-space diagrams \citep{2011MNRAS.416.2882M,2013MNRAS.431.2307O,2016MNRAS.463.3083O,2019MNRAS.484.1702P}, which use both projected group-centric velocity and projected group-centric radius.  The range of the positions of satellites in groups can be defined by a caustic profile \citep{2019MNRAS.484.1702P}: $K = |\Delta V|/\sigma \times R/R_{180}$, where |$\Delta$V| is the peculiar line-of-sight velocity and $\sigma$ is the velocity dispersion along the line of sight, which is obtained by computing the standard unbiased variance of the line-of-sight velocities. The value of $K$ will change with the infall time of the satellites, with ancient infallers having smaller values compared to recent infallers \citep{2019MNRAS.484.1702P}. 

\begin{figure}
    \centering
    \includegraphics[width=8cm]{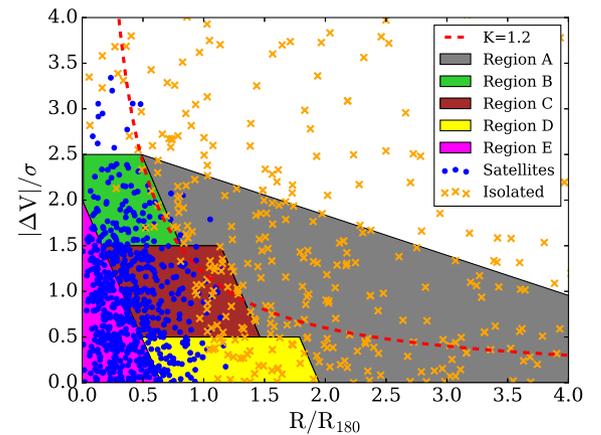}
    \caption{The phase-space diagram for the satellites (blue points) and isolated galaxies (orange crosses) in our sample. The red dashed line refers to the the caustic profile with $K = 1.2$. The colored five regions are given by \citet{2017ApJ...843..128R}, based on the time since infall.}
    \label{phase_diagram}
\end{figure}

Using cosmological hydrodynamic N-body simulations of groups and clusters, \citet{2017ApJ...843..128R} investigate how the information on environmental effects acting in groups and clusters
 can be extracted by the knowledge of locations in phase-space coordinates. They show that galaxies tend to follow a typical path in phase-space when falling into the groups. Based on the time since infall, they divide the projected phase-space diagram into five regions. These regions are chosen to maximize the proportion of galaxies that belong to a specific time since they fall into the groups.

Figure~\ref{phase_diagram} shows the positions of our satellites in ($R$, |$\Delta$V|) projected phase space, where $R$ refers to the projected group-centric radius and |$\Delta$V| is the peculiar line-of-sight velocity of each satellite, determined as the absolute value of the difference between its line-of-sight velocity and the average line-of-sight velocity of all satellites in the same group. All the velocities here are derived from spectroscopic redshifts. In Figure~\ref{phase_diagram}, the satellites and isolated galaxies are labeled by blue points and orange crosses. The projected group-centric radius for isolated galaxies refers to the projected distance to groups which are closest to them. Figure~\ref{phase_diagram} shows that the isolated galaxies and satellites are located in different areas in phase space, with most of the satellites located inside the $|\Delta V|/\sigma \times R/R_{180}$ = 1.2 profile (red dashed line in Figure~\ref{phase_diagram}). 

We also show the infall regions given by \citet{2017ApJ...843..128R} in Figure~\ref{phase_diagram}. We scale the phase-space plot in R$_{180}$ with R$_{180}$ $\sim$ 0.77 R$_{vir}$, assuming a Navarro, Frenk and White
\citep[NFW;][]{1997ApJ...490..493N} profile with concentration parameter c=4. The dominant galaxy populations in regions A, B, D and E are first infallers, recent infallers, intermediate infallers and ancient infallers, respectively. Region C is a mixing area with each population taking similar fraction. They find the galaxies follow the path in order of A, B, C, D, E, as galaxies settle into groups potentials.

\begin{figure}
    \centering
    \includegraphics[width=8cm]{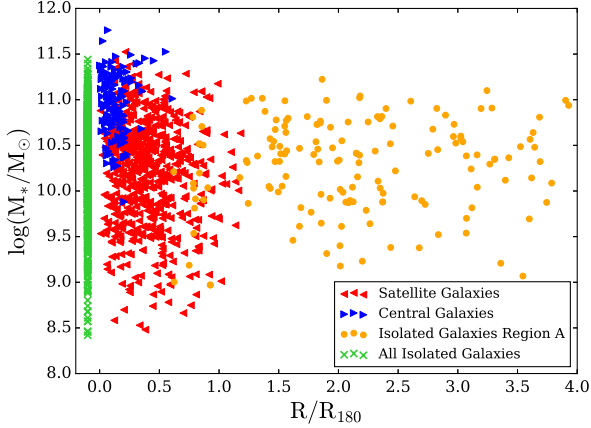}
    \caption{Stellar mass as a function of normalised projected group-centric radius for all galaxies in our sample. The centrals in groups, isolated galaxies and satellites are marked with blue right-triangles, green crosses and red left-triangles, respectively. Isolated galaxies are shifted from zero projected radius for clarity. The orange points are isolated galaxies located within Region A (see Figure~\ref{phase_diagram}) of a neighbouring group, and the group-centric radius is the distance to their nearest neighbouring groups.}
    \label{stellarmass_distance_diagram_allgalaxies}
\end{figure}

We use the bound of region A (grey) to identify the infalling isolated galaxies in our sample and to extend our measurements to larger projected radii.  Figure~\ref{stellarmass_distance_diagram_allgalaxies} shows stellar mass as a function of normalised projected group-centric radius for all galaxies (centrals in groups, isolated galaxies and satellites). 
Isolated galaxies within region A are included, with group-centric radius being the distance to the neighbouring group centre. 
In Figure~\ref{stellarmass_distance_diagram_allgalaxies}, we separately show all the isolated galaxies (not limited by the criterion above) in our sample at zero projected radius (the centre of their groups are themselves).

\begin{figure*}
\begin{multicols}{3}
\includegraphics[width=6.cm,height=4.9cm]{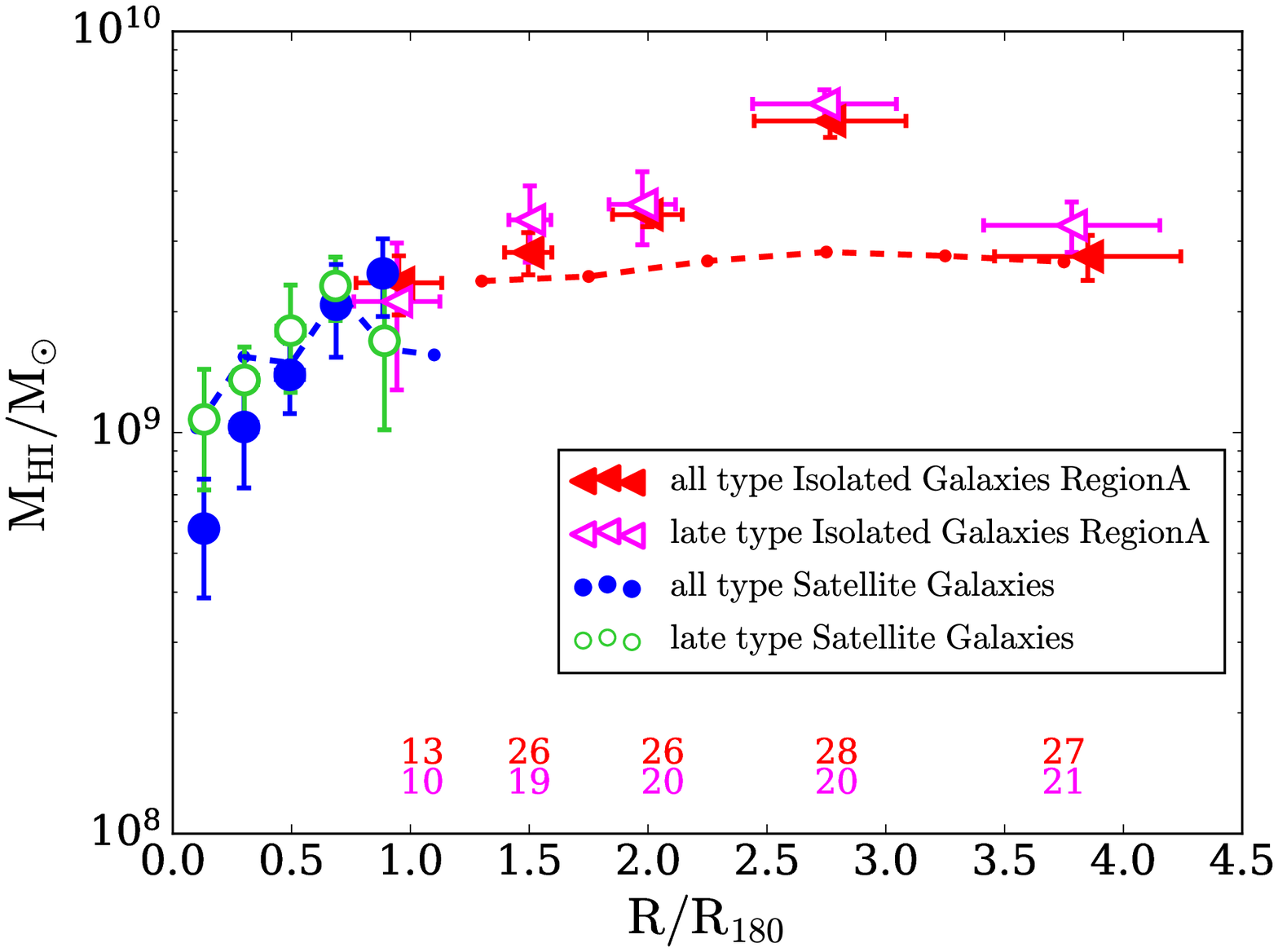}\par
\hspace*{-0.3cm}
\includegraphics[width=6.cm,height=4.9cm]{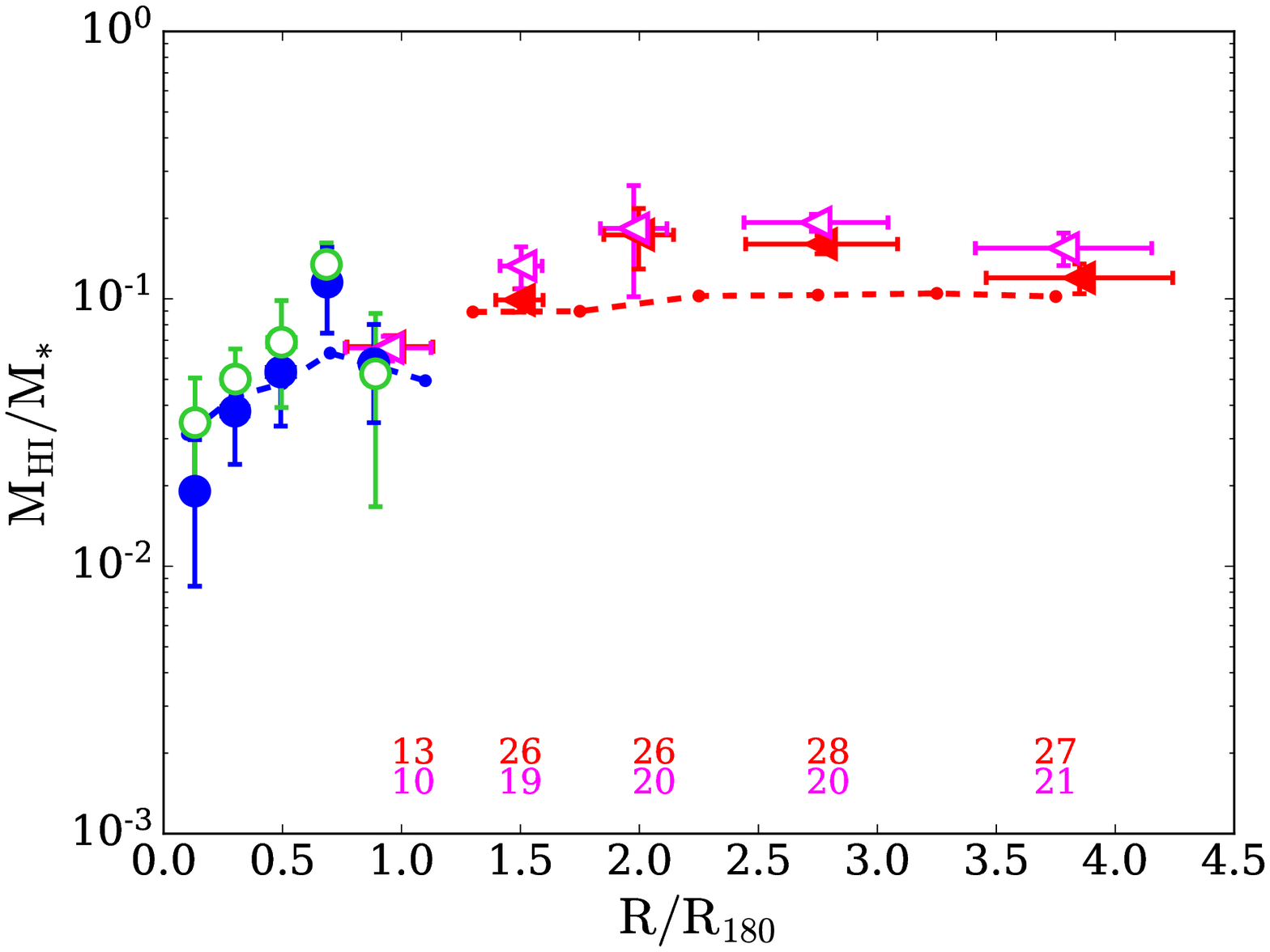}\par
\hspace*{-0.3cm}
\includegraphics[width=6.cm,height=4.9cm]{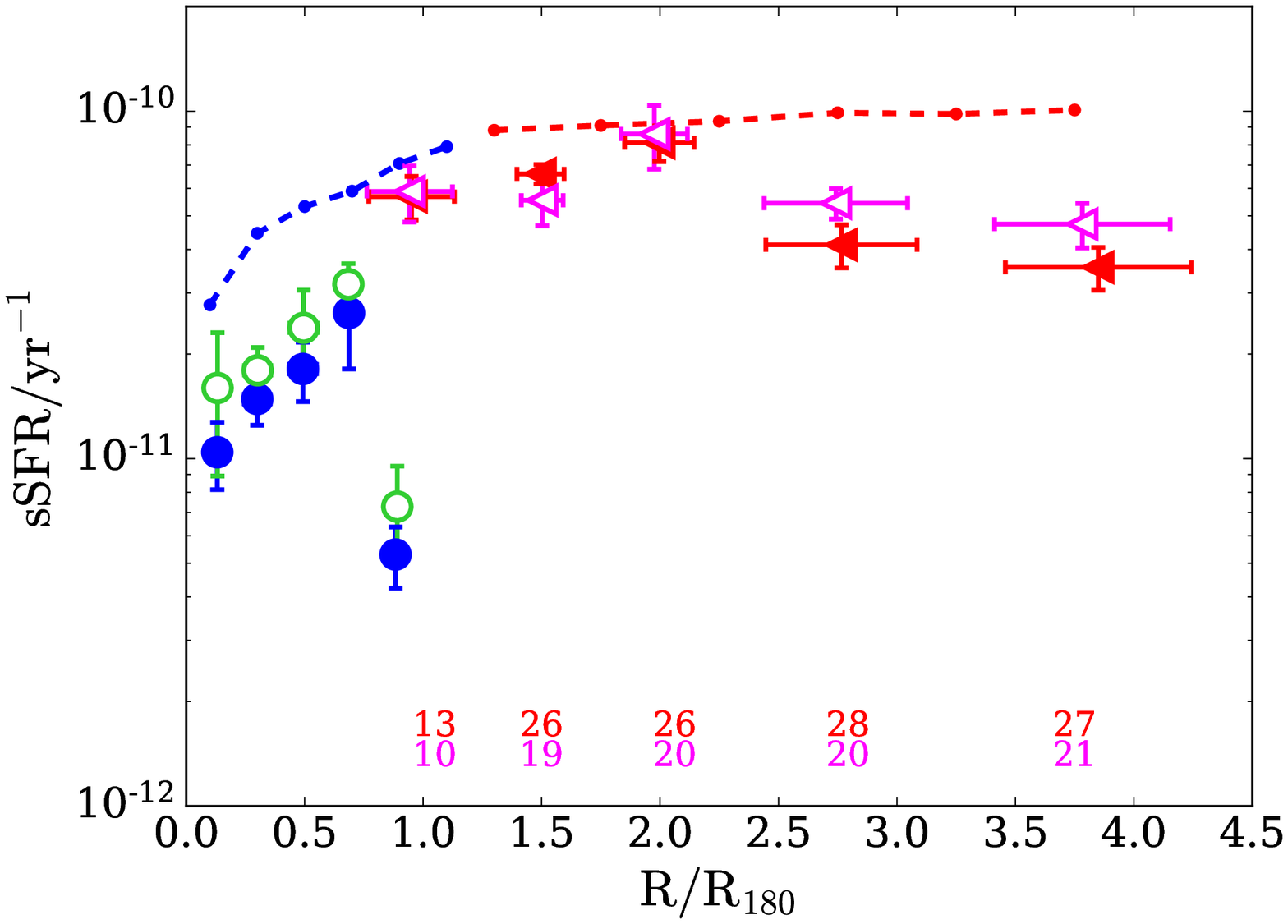}\par
\end{multicols}
\caption{Same as Figure~\ref{distance_HIcontent_satellite}, but for satellite galaxies (circle points; see Fig.~\ref{distance_HIcontent_satellite}) and infalling isolated galaxies (left-pointing triangles) in the range $10^{10.0} < M_{\ast} \leqslant 10^{11.5}$M$_{\odot}$ as a function of normalised projected group-centric radius, extended out to greater radii. The results from the all-type and the late-type galaxies are labeled as filled and open points, respectively. For the isolated galaxies, the group-centric radius correspond to the distance from the galaxy to the centre of its nearest group. The corresponding values are presented in Table~\ref{distance_HIcontent_extend_isolates_tab} and Table~\ref{distance_HIcontent_satellite_tab}. Each panel also overplots the measurement from {\sc Shark} simulation with red and blue dashed line corresponding to the all-type satellite galaxies and infalling isolated galaxies.}
\label{distance_HIcontent_extend_isolates}
\end{figure*}

Now that we have a radial distance for each isolated galaxy close to a group, we stack them in bins of projected group-centric radius. As before, we only use galaxies in the range $10^{10.0} < M_{\ast} \leqslant 10^{11.5}$M$_{\odot}$. The results in Figure~\ref{distance_HIcontent_extend_isolates} and Table~\ref{distance_HIcontent_extend_isolates_tab} show that isolated galaxies near neighboring groups lack \hi\ relative to isolated galaxies farther away from neighbouring groups. The \hi\ gas fraction increases with normalised projected group-centric radius until $R$ $\sim$ 2.0 $R_{180}$. The sSFR also increases with the group-centric radius to $R$ $\sim$ 2.0 $R_{180}$. For comparison, we also reproduce here the \hi\ properties and sSFR stacking results of satellite galaxies from Figure~\ref{distance_HIcontent_satellite},
labeled as circle points. The \hi\ properties and sSFR vs. group centric radius relations for isolated galaxies can be well connected to those for satellite galaxies. The increasing trend with radius for isolated galaxies seems to be the continuation of the trend for satellites. This suggests that \hi\ gas loss starts well before a galaxy reaches R$_{180}$ of a group and formally becomes a satellite. 

For completeness, we also measured the \hi\ content of galaxies as a function of local 3D density. We found that for centrals, satellites and isolated galaxies, the \hi\ mass decreases with increasing 3d density, which is consistent with the stacking results as a function of distance. This is because high local 3D densities always correspond to small group-centric radii. 

The stacked mass spectra for isolated galaxies are shown in Appendix (Figure~\ref{stacked_spectra_isolate}).

\begin{table*}
 \caption{Some basic statistics information and the stacking results of \hi\ properties as a function of group-centric radius, for the all-type and the late-type isolated galaxies at 10$^{10.0}$M$_{\odot}$ < $M_{\ast}$ $\leqslant$ 10$^{11.5}$M$_{\odot}$, extended to larger radii. We illustrate the results in Figure~\ref{distance_HIcontent_extend_isolates}.}
 \label{distance_HIcontent_extend_isolates_tab}
  \begin{tabular}{lccccccc}
   \hline
   Radial bins & Radius & Number of galaxies & $r_{50}/r_{90}$ & $\langle M_{\ast}\rangle$ & $\langle M_{\rm \hi}\rangle$ & $\langle M_{\rm \hi}/M_{\ast}\rangle$ & $\langle sSFR\rangle$\\
    ($R_{180}$) & ($R_{180}$) & & & $(10^{10} h_{70}^{-2}$ M$_{\odot})$ & $(10^{9} h_{70}^{-2}$ M$_{\odot})$ & & (10$^{-11}$/yr$^{-1}$)\\
    \hline
     all-type isolated galaxies\\
   (0.75,1.25] & 1.04 $\pm$ 0.14 & 13 & 0.41 $\pm$ 0.06 & 4.1 $\pm$ 0.3 & 2.3 $\pm$ 0.5 & 0.07 $\pm$ 0.01 & 5.7 $\pm$ 0.9\\
   (1.25,1.75] & 1.50 $\pm$ 0.12 & 26 & 0.39 $\pm$ 0.07 &  4.3 $\pm$ 0.2 & 2.8 $\pm$ 0.3 & 0.10 $\pm$ 0.01 & 6.6 $\pm$ 0.4\\
   (1.75,2.25] & 1.98 $\pm$ 0.15 & 26 & 0.39 $\pm$ 0.06 &  4.4 $\pm$ 0.4 & 3.5 $\pm$ 0.3 & 0.17 $\pm$ 0.04 & 8.1 $\pm$ 0.9\\
   (2.25,3.25] & 2.76 $\pm$ 0.32 & 28 & 0.38 $\pm$ 0.06 &  4.6 $\pm$ 0.3 & 6.0 $\pm$ 0.6 & 0.16 $\pm$ 0.01 & 4.1 $\pm$ 0.8\\
   (3.25,4.45] & 3.85 $\pm$ 0.39 & 27 & 0.38 $\pm$ 0.06 &  4.2 $\pm$ 0.3 & 2.8 $\pm$ 0.4 & 0.12 $\pm$ 0.02 & 3.6 $\pm$ 0.4\\
   \hline
     late-type isolated galaxies\\
   (0.75,1.25] & 1.04 $\pm$ 0.14 & 10 & 0.43 $\pm$ 0.03 &  3.1 $\pm$ 0.2 & 2.1 $\pm$ 0.3 & 0.06 $\pm$ 0.01 & 5.9 $\pm$ 1.0\\
   (1.25,1.75] & 1.50 $\pm$ 0.12 & 19 & 0.42 $\pm$ 0.05 &  3.6 $\pm$ 0.3 & 3.4 $\pm$ 0.9 & 0.13 $\pm$ 0.02 & 5.5 $\pm$ 0.5\\
   (1.75,2.25] & 1.98 $\pm$ 0.15 & 20 & 0.41 $\pm$ 0.05 &  3.6 $\pm$ 0.5 & 3.7 $\pm$ 0.8 & 0.18 $\pm$ 0.08 & 8.6 $\pm$ 1.6\\
   (2.25,3.25] & 2.76 $\pm$ 0.32 & 20 & 0.41 $\pm$ 0.04 &  3.9 $\pm$ 0.3 & 6.6 $\pm$ 0.5 & 0.19 $\pm$ 0.01 & 5.4 $\pm$ 0.7\\
   (3.25,4.45] & 3.85 $\pm$ 0.39 & 21 & 0.40 $\pm$ 0.05 &  3.6 $\pm$ 0.3 & 3.3 $\pm$ 0.4 & 0.15 $\pm$ 0.02 & 4.7 $\pm$ 0.7\\
   \hline
  \end{tabular}
\end{table*}

\vspace*{-0.5cm}
\section{Discussion}
\label{sec:discussion}

\citet{2016ApJ...824..110O} presented the \hi\ content of galaxies measured by the 70$\%$ complete ALFALFA survey and study the \hi distribution in nearby groups and clusters. They compared the \hi\ content in galaxies at fixed stellar mass and galaxy type in the centres of groups and clusters with the \hi\ content in galaxies in control regions out to 4 Mpc surrounding each group or cluster. They found that at fixed stellar mass, the late-type galaxies in the centres of groups lack \hi\ compared with galaxies in the outer control region. This is consistent with our results that at fixed stellar mass the satellites in the centres of groups lack \hi\ relative to those at larger radii.

In Section~\ref{sec:Environmental dependence}, we compared our radial trends for low-mass and high-mass groups. Although the trend is more significant in high-mass groups, there is still  an increase of \hi\ mass with increasing distance from the group centre in low-mass groups with M$_{\rm halo}$ below $10^{13.5}h^{-1}$M$_{\odot}$, well before galaxies reach the cluster environment. This is consistent with \citet{2016ApJ...824..110O} and \citet{2017MNRAS.466.1275B}, indicating existence of \hi\ removal in isolated groups. The same conclusions are reached if we bin galaxies by projected densities, instead of group halo mass. However, for low densities the decrease of \hi\ content in the center of groups practically disappears.

In order to find the best predictors of galaxy properties, \citet{2016ApJ...824..110O} ran regressions against six environment variables (group-centric radius $r$, normalized group-centric radius $r/R_{200}$, density $\Sigma$, group mass $M_{200}$, halo mass in the Yang catalog, and central/ satellite status in the Yang catalog). By comparing the standardized slopes from regressions for log stellar mass, g-i color, log \hi\ mass, and \hi\ deficiency for blue cloud galaxies as a function of six different environment variables, they found that local density is the most effective predictor, while $r/R_{200}$ and group-centric radius $r$ are similarly less effective, followed by group size and halo mass. However, the opposite conclusion was reached by \citet{2017MNRAS.466.1275B}, who stacked the \hi\ spectra of 10,600 satellite galaxies measured by the ALFALFA survey to investigate environment-driven gas depletion in satellite galaxies. \citet{2017MNRAS.466.1275B} showed that gas content is depleted with increasing fixed aperture and nearest neighbour densities, but that halo mass is the most dominant environmental driver of \hi\ removal in satellites. Specifically, when one fixes density and alters the halo mass, differences are larger than when density is changed at fixed halo mass. Besides, it is shown that at fixed sSFR gas fraction decreases more significantly with halo mass than with density. The conflicting results of which one of local density and halo mass can more effectively drive environmental \hi\ removal are most likely due to different sample selection. \citet{2016ApJ...824..110O} worked only with ALFALFA detections, while \citet{2017MNRAS.466.1275B} used staking. So, \citet{2016ApJ...824..110O} focused on gas-rich galaxies for which the environment has just started affecting their evolution, while \citet{2017MNRAS.466.1275B} covered the entire range of gas fraction.

Following what done in \citet{2020MNRAS.493.1587H}, we compare our results with the prediction from the {\sc Shark} \citep{2018MNRAS.481.3573L} semi-analytic model. We construct a lightcone with an area of $\sim 6900$deg$^{2}$ and redshift range of $z=0-0.1$, containing all the galaxies with $M_{\star} \geq 10^{5} M_{\odot}$ (see \citealt{Chauhan2019} for details on how lightcones are constructed). Using the {\sc Shark} lightcone and the method described in Section~\ref{sec:script}, we stack the \hi\ mass, \hi\ mass fraction and sSFR from {\sc Shark} galaxies with stellar masses $10^{10.0} < M_{\ast}/$M$_{\odot} \leqslant 10^{11.5}$ and apparent $r$-band magnitude $M_{r} < 17.7$~mag. For the stacking of isolated galaxies in {\sc Shark}, we only use the infalling galaxies in Region A (see Section~\ref{sec:larger_distance}). The results are presented in Figure~\ref{distance_HIcontent_extend_isolates}, as dashed lines.

Overall, {\sc Shark} qualitatively matches the radial trends of satellite galaxies (red points and lines) for both \hi\ mass and \hi\ mass fraction, with the only potential difference being a milder decrease toward the centre of groups, compared to our observational results. This is not surprising, given that the model does not include stripping of the cold gas phase. Indeed, \citet{2020MNRAS.493.1587H} already highlighted the limitation of  {\sc Shark} in reproducing the properties of satellite galaxies. 

However, the most striking difference between the model and our observations is in the sSFR radial trends, where {\sc Shark} produces sSFRs a factor of $\sim$2-3 systematically higher than what measured using our sample. The origin of this difference is unclear. Potential candidates are too high molecular gas masses and/or star formation efficiencies, suggesting that the lack of implementation of cold gas stripping alone cannot fully explain the differences highlighted here, as well as those presented in \citet{2020MNRAS.493.1587H}.

Because of the limited size of our sample, we can't firmly determine which quantity between local density and halo mass is more closely associated to environment driven gas depletion. From the evidence presented here, it is possible to take our analysis a step further and quantify the \hi\ properties vs. group-centric radius relations at fixed stellar mass as a function of halo mass and density, only when a large enough galaxy sample can be accessed. Next-generation Square Kilometre Array (SKA) precursor facilities
such as the Australian SKA Pathfinder (ASKAP) \citep{2008ExA....22..151J,2009pra..confE..15M}, MeerKAT \citep{2012IAUS..284..496H}, Five-hundred-meter Aperture Spherical radio Telescope (FAST) \citep{2011IJMPD..20..989N,2008MNRAS.383..150D,2020MNRAS.493.5854H} and WSRT/Aperture Tile in Focus (APERTIF) \citep{2009wska.confE..70O}
will enable large-area surveys to significant depths. It will very soon be possible to extend our measurements to significantly larger samples, making it feasible to further quantify the contribution of different mechanisms to the \hi\ content increasing trend on radius, and potentially extend this to higher redshift. 

Nevertheless, our results clearly highlight how the effect of environment on the gas content of galaxies may start well before they enter into the satellite phase, suggesting a potential role of pre-processing in the evolution of group and cluster galaxies (see also the recent work by \citet{2021arXiv210104389C}). While statistically significant, it is important to remember that the effect of pre-processing does not affect \hi\ content more than a factor of $\leq$2, implying that when galaxies formally become satellites they are still gas-normal. In other words, the environmental processes making galaxies \hi\ deficient, and ultimately passive, appear to be generally confined in the inner parts of groups and more efficient at high halo masses and projected galaxy density.

\vspace*{-0.5cm}
\section{Summary}
\label{sec:summary}
In this work we utilize an interferometric spectral stacking technique to study the distribution of \hi\ content of galaxies and show that the galaxies in the centres of groups lack \hi\ at fixed stellar mass and morphology relative to satellites in the group outskirts. 

The data come from a 351-hr WSRT \hi\ survey covering $\sim 35$ deg$^2$ of the SDSS sky. After cross-matching with Yang's group catalogue, 1793 galaxies with SDSS redshifts in the range $0.01 < z < 0.11$ are available for stacking. 120 isolated galaxies and 457 satellites with stellar masses $10^{10.0} < M_{\ast}/$M$_{\odot} \leqslant 10^{11.5}$ are used to quantify the atomic hydrogen content of galaxies as a function of group-centric radius.

We find that the \hi\ content of satellites monotonically decreases getting closer to the center of groups. A similar trend is found for the sSFR. We test the trends against two environment variables, namely halo mass and local density. This shows that, for both high-mass and low-mass groups, galaxies in the central regions have smaller \hi\ content than those in the outer regions, with the trends being more evident in high-mass groups. For galaxies in high local density regions, the \hi\ content increases with group-centric radius. However, no clear relation between \hi\ content and radius is apparent for galaxies in regions of low local density.
Excitingly, we show that these radial trends continue at distances larger than $R_{180}$, once isolated galaxies in the infalling regions of groups are included. This provides tantalising evidence that gas removal may start well before the crossing of the virial radius in groups and clusters. We also measure these trends in the late-type subsample to quantify the influence of morphology on our findings, and obtain similar results.

We compare our measurements with the semi-analytic model {\sc Shark} and find that, although the model reproduces the general decrease of \hi\ mass with decreasing group-centric radius, it still fails in reproducing both \hi\ and sSFR properties of our sample, simultaneously.

\vspace*{-0.5cm}
\section{Data Availability}
The radio data analysed in this work are from WSRT and firstly reported in \citet{2015A&A...580A..43G}. The derived data can be accessed by sending request to the corresponding authors of this paper. The corresponding optical catalog is from SDSS \citep{2000AJ....120.1579Y} Data Release 7.

\vspace*{-0.5cm}
\section{Acknowledgements}
We thank the referee for a constructive report that helped improving our paper. The WSRT is operated by ASTRON (Netherlands Foundation for Research in Astronomy) with support from the Netherlands Foundation for Scientific Research (NWO). We acknowledge the use of Miriad software in our data analysis (http://www.atnf.csiro.au/computing/software/miriad/). This research made use of the Sloan Digital Sky Survey archive. The full acknowledgment can be found at http://www.sdss.org. LC is the recipient of an Australian Research Council Future Fellowship (FT180100066) funded by the Australian Government. Parts of this research were supported by the Australian Research Council Centre of Excellence for All Sky Astrophysics in 3 Dimensions (ASTRO 3D), through project number CE170100013. Wenkai Hu is supported from the European Research Council (ERC) under the European Union's Horizon 2020 research and innovation programme (project CONCERTO, grant agreement No 788212) and from the Excellence Initiative of Aix-Marseille University-A*Midex, a French "Investissements d'Avenir" programme.

\bibliographystyle{mnras}
\bibliography{group.bib}

\vspace*{-0.5cm}
\appendix
\section{Stacked Spectra}
\label{sec:StackedSpectra}
\begin{figure}
\begin{multicols}{2}
\includegraphics[width=4.5cm,height=3.75cm]{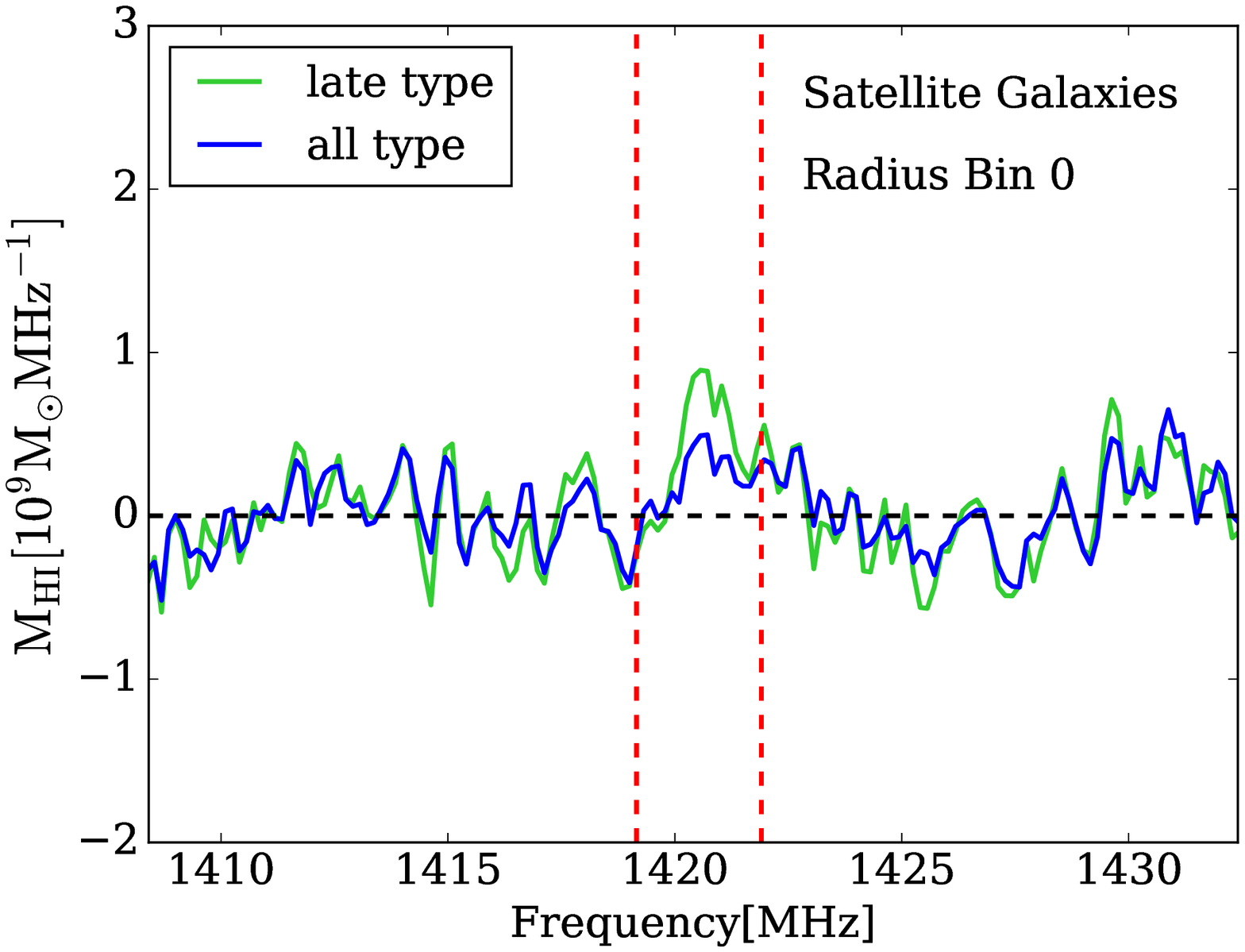}\par
\hspace*{-0.3cm}
\includegraphics[width=4.5cm,height=3.75cm]{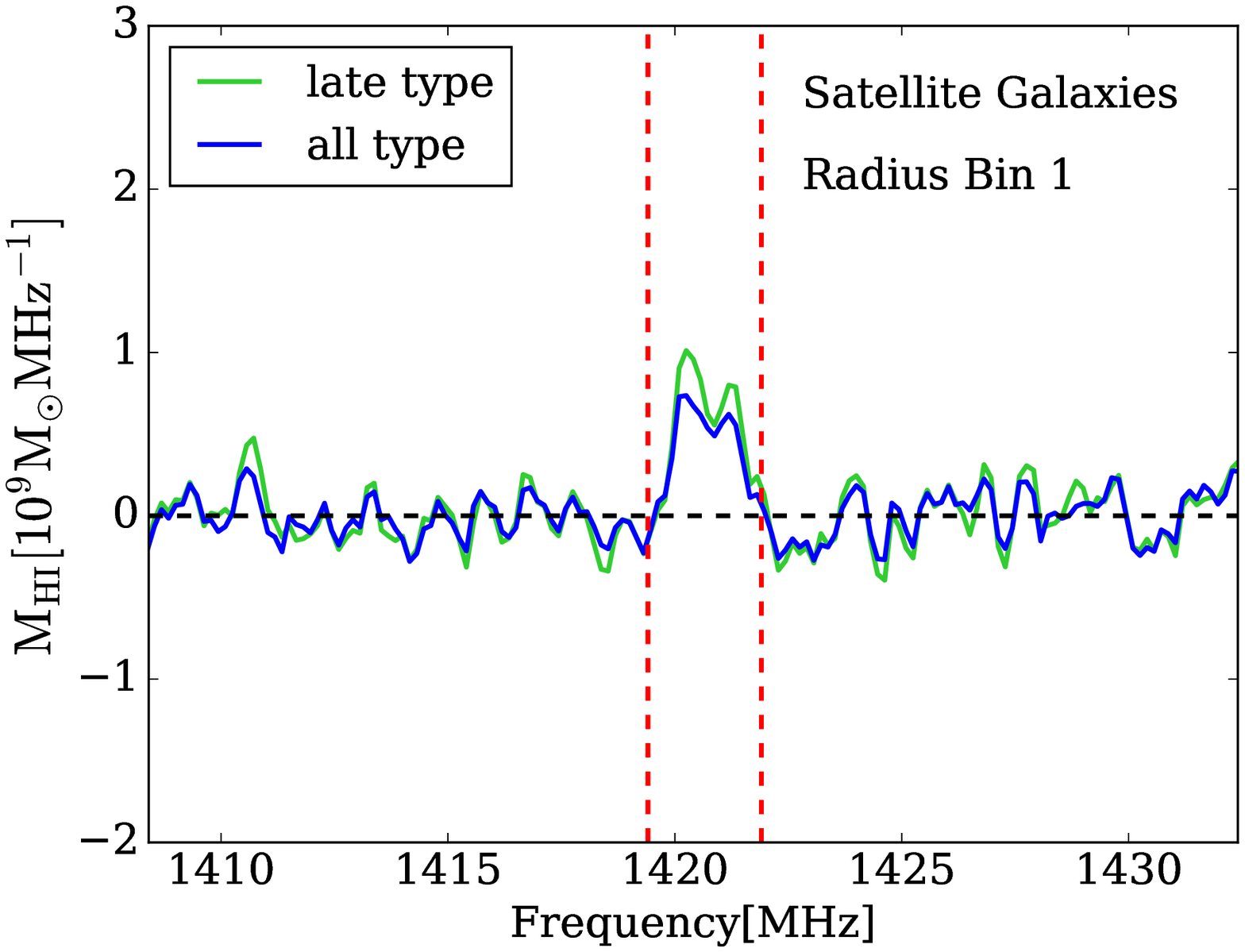}\par
\end{multicols}
\vspace*{-1.05cm}
\begin{multicols}{2}
\includegraphics[width=4.5cm,height=3.75cm]{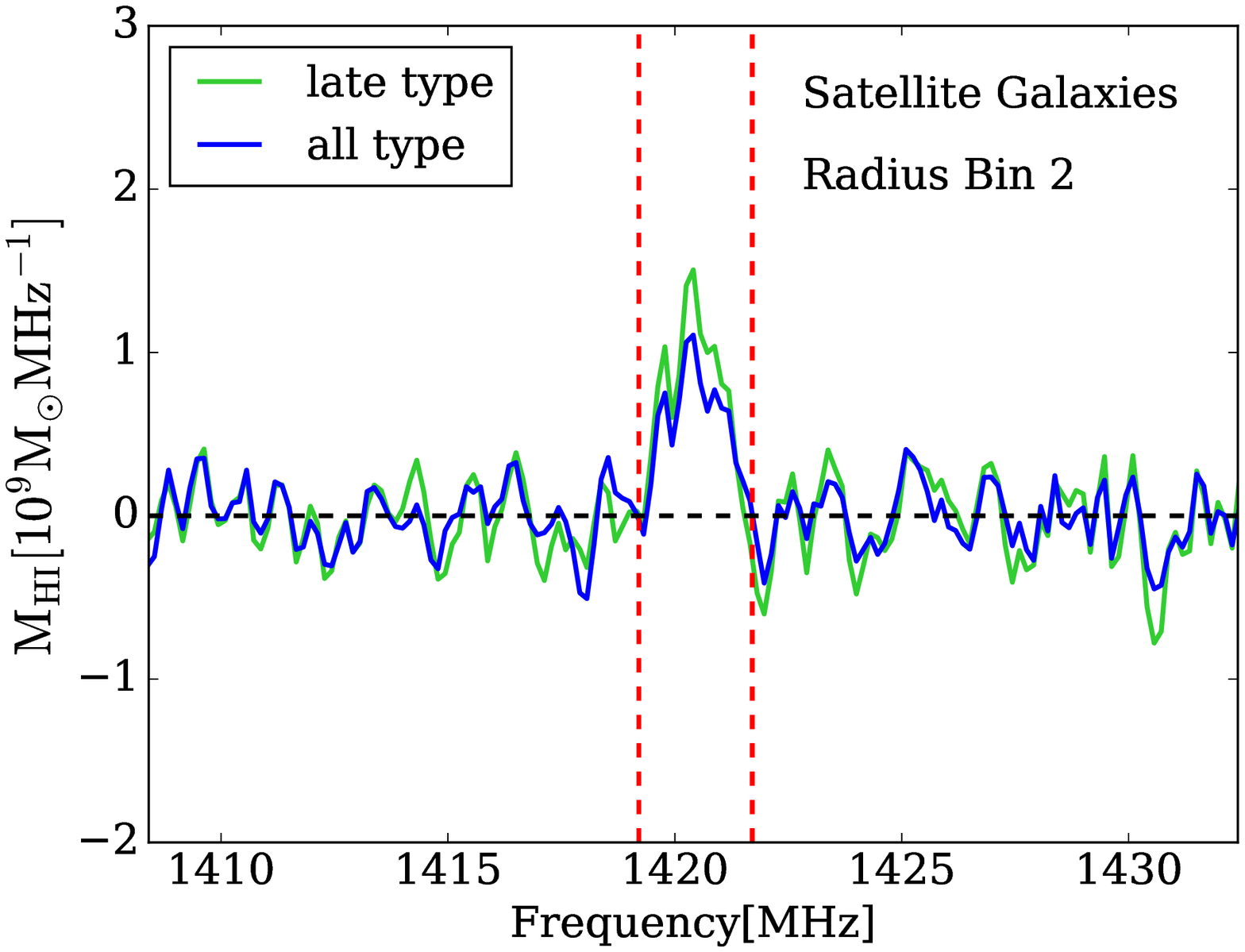}\par
\hspace*{-0.3cm}
\includegraphics[width=4.5cm,height=3.75cm]{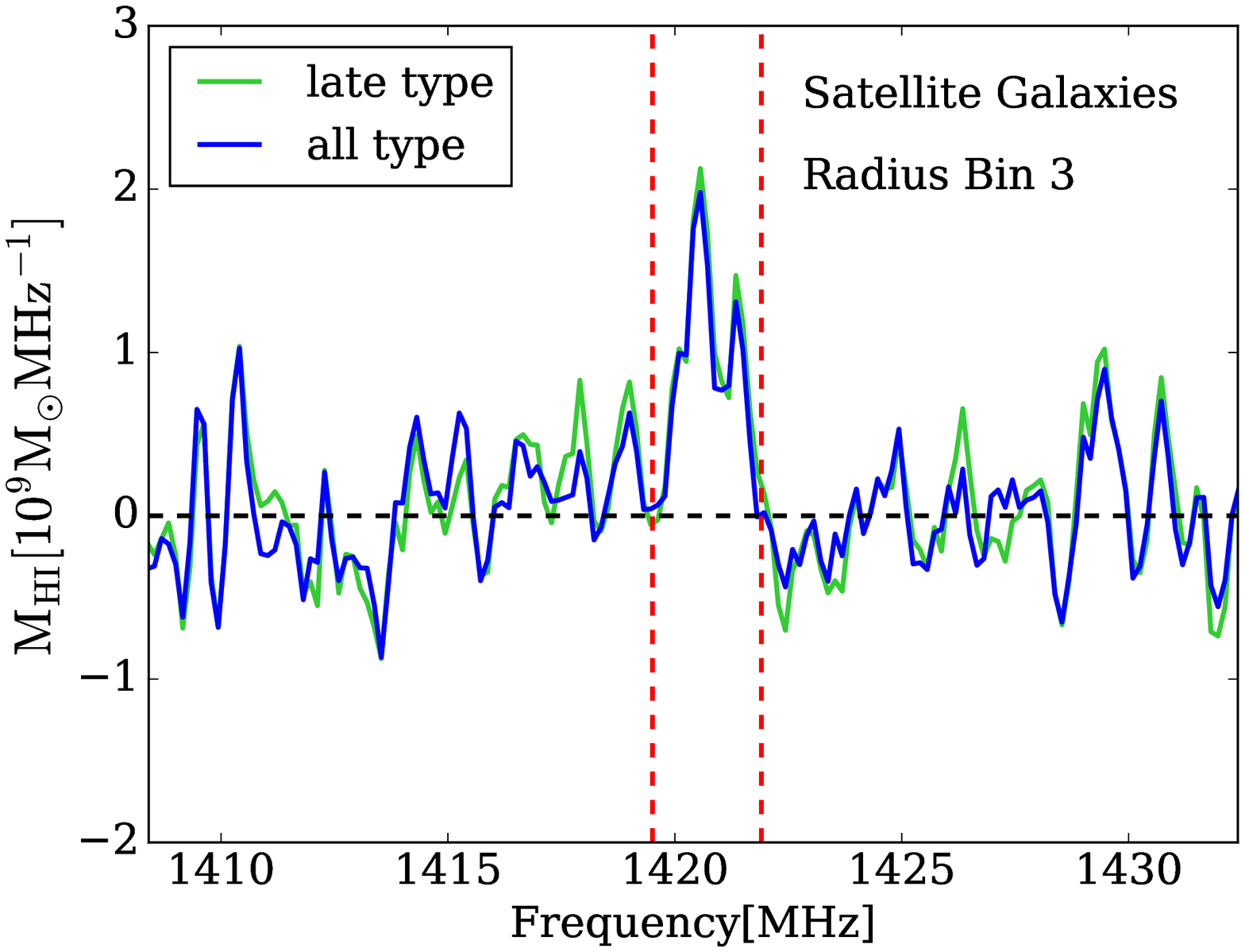}\par
\end{multicols}
\vspace*{-1.05cm}
\begin{multicols}{2}
\includegraphics[width=4.5cm,height=3.75cm]{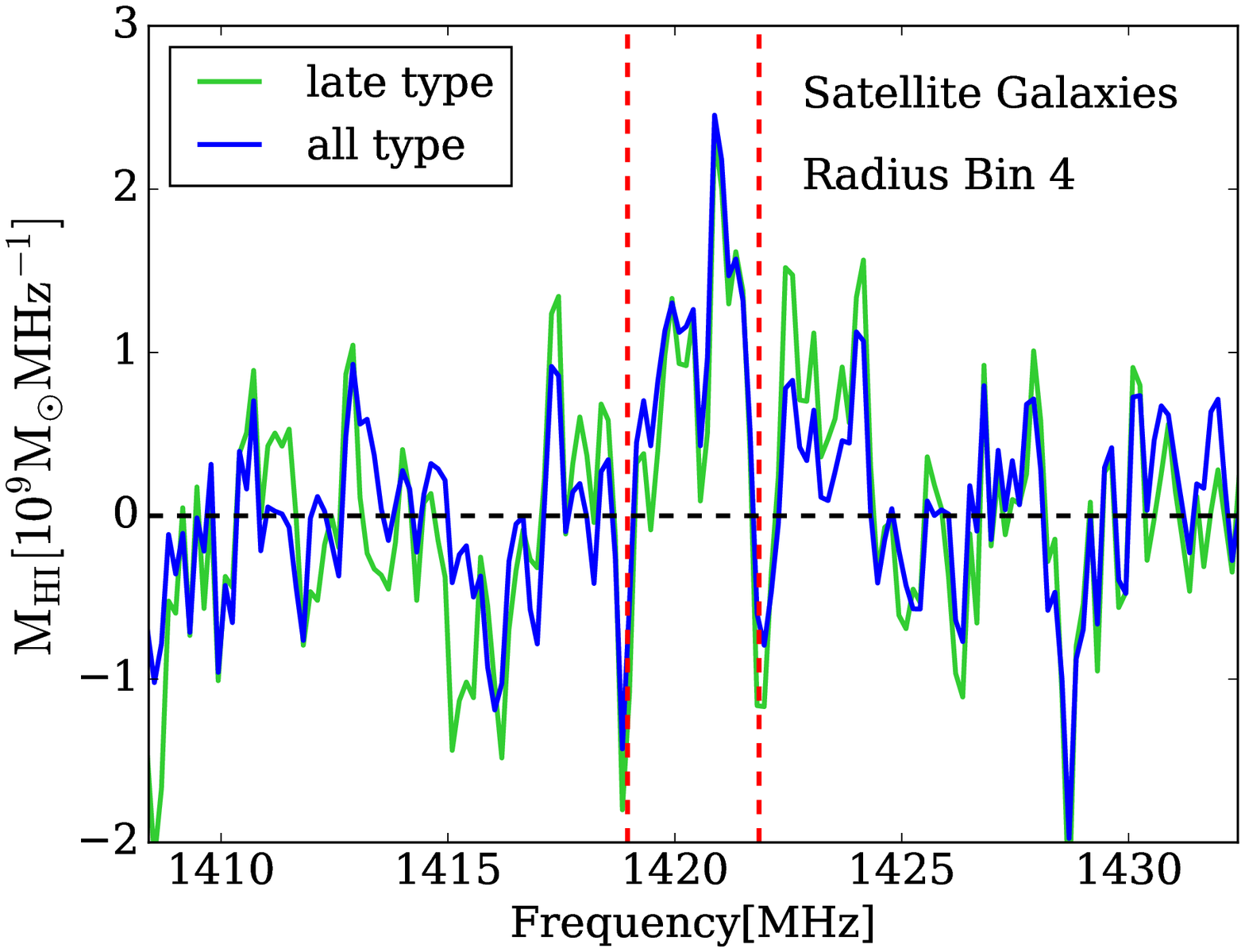}\par
\end{multicols}
\vspace*{-0.8cm}
\caption{The stacked mass spectra for satellite galaxies.}
\label{stacked_spectra_satellite}
\end{figure}

\begin{figure}
\begin{multicols}{2}
\includegraphics[width=4.5cm,height=3.75cm]{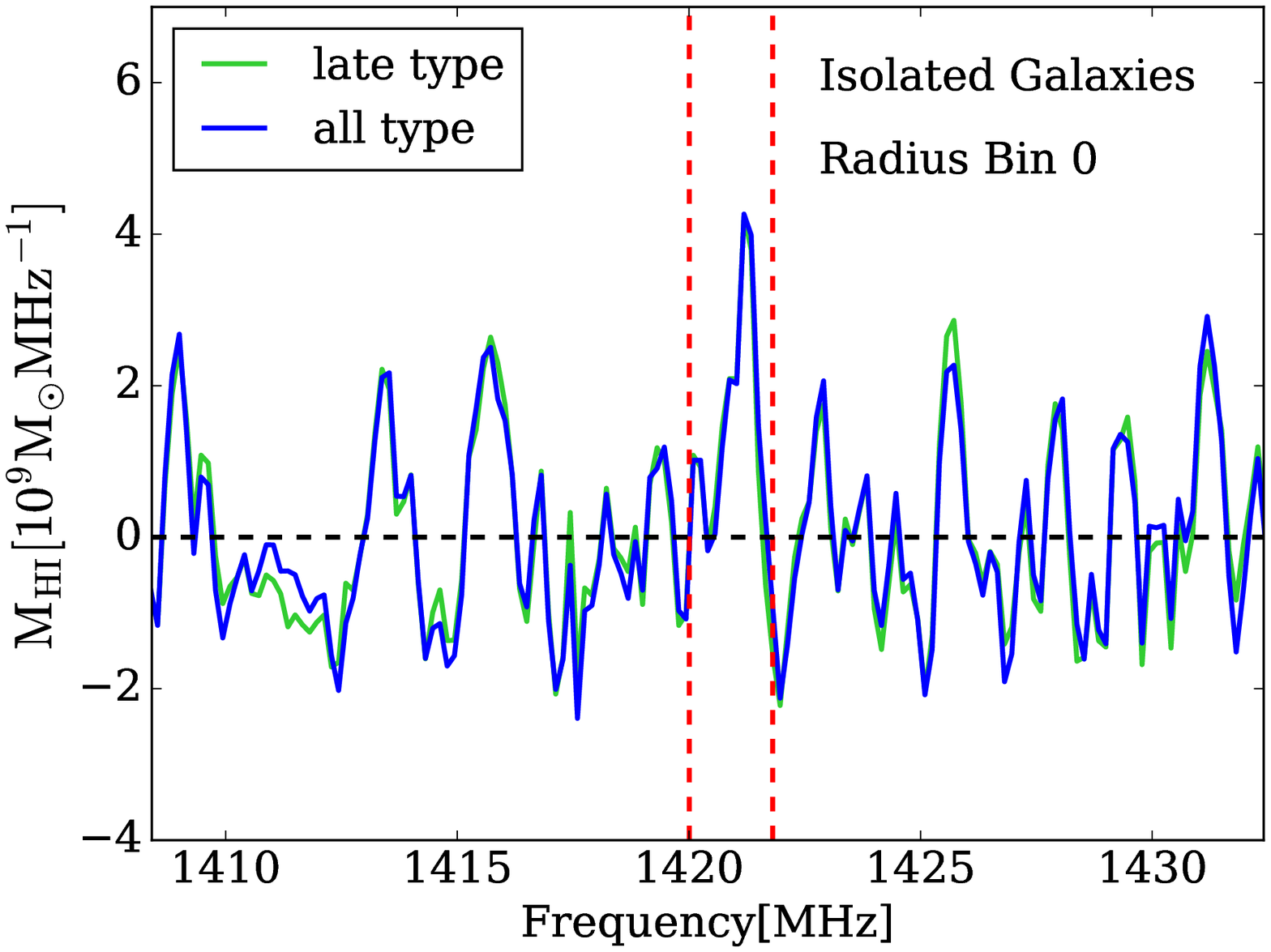}\par
\hspace*{-0.3cm}
\includegraphics[width=4.5cm,height=3.75cm]{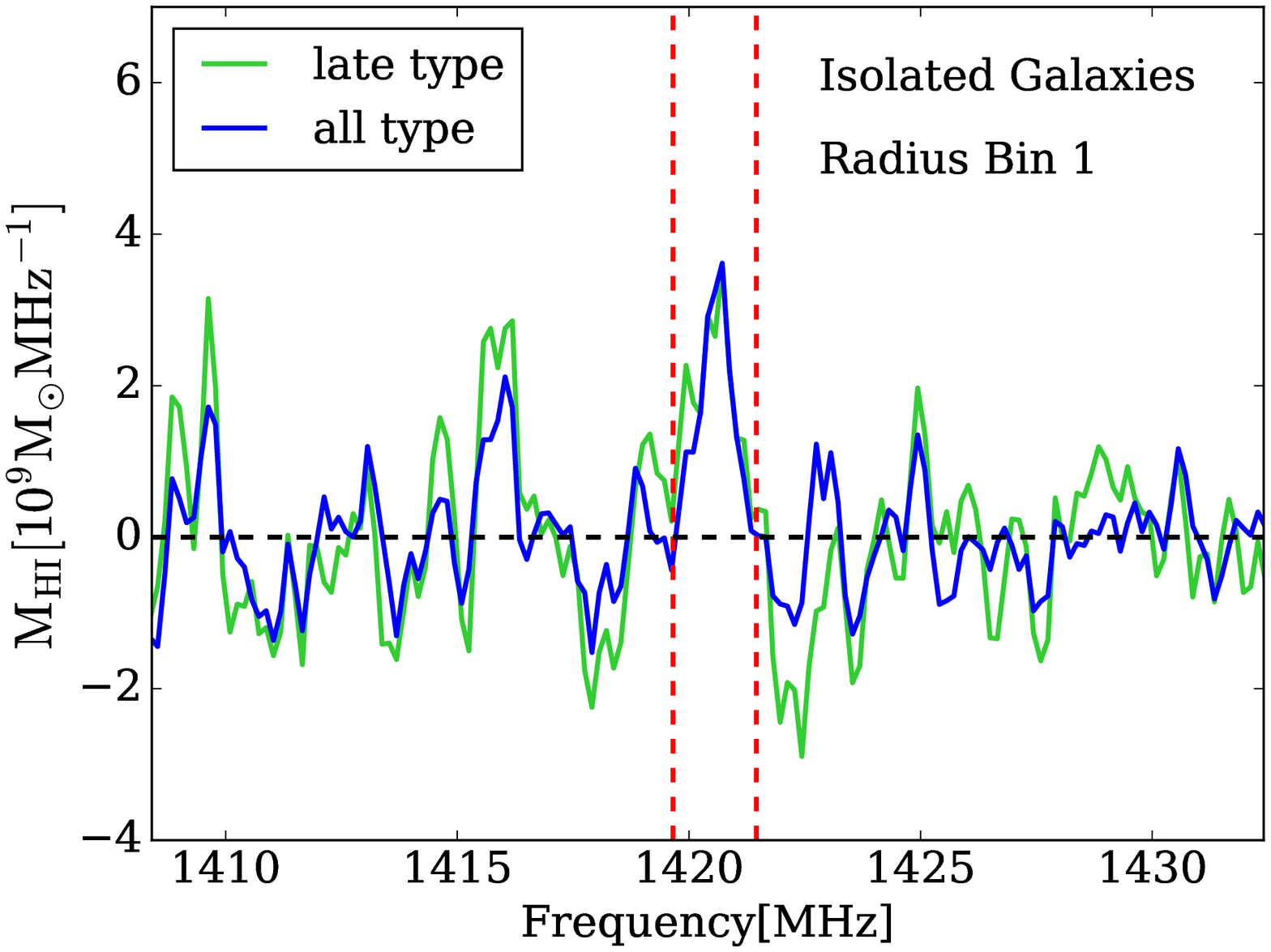}\par
\end{multicols}
\vspace*{-1.05cm}
\begin{multicols}{2}
\includegraphics[width=4.5cm,height=3.75cm]{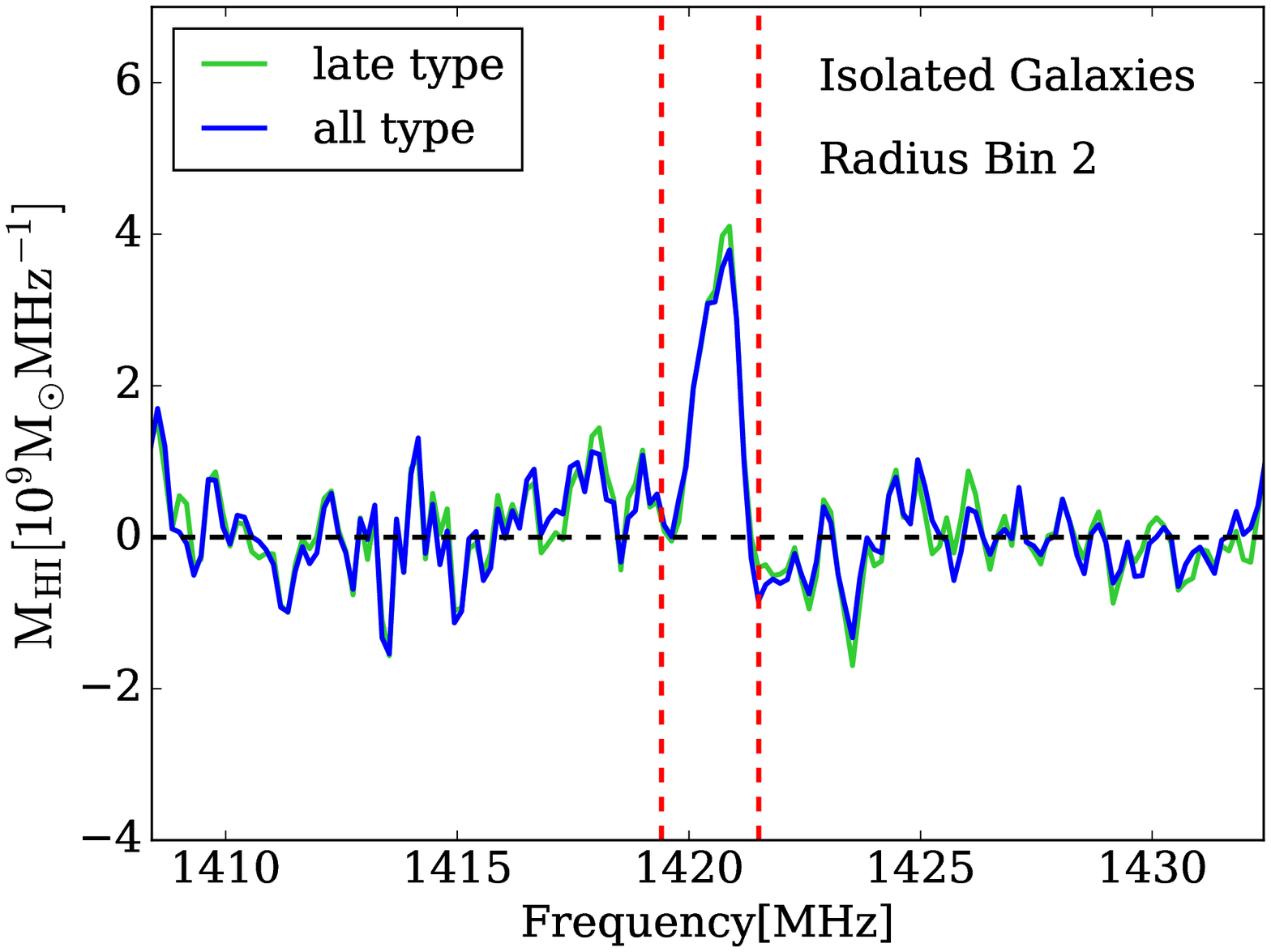}\par
\hspace*{-0.3cm}
\includegraphics[width=4.5cm,height=3.75cm]{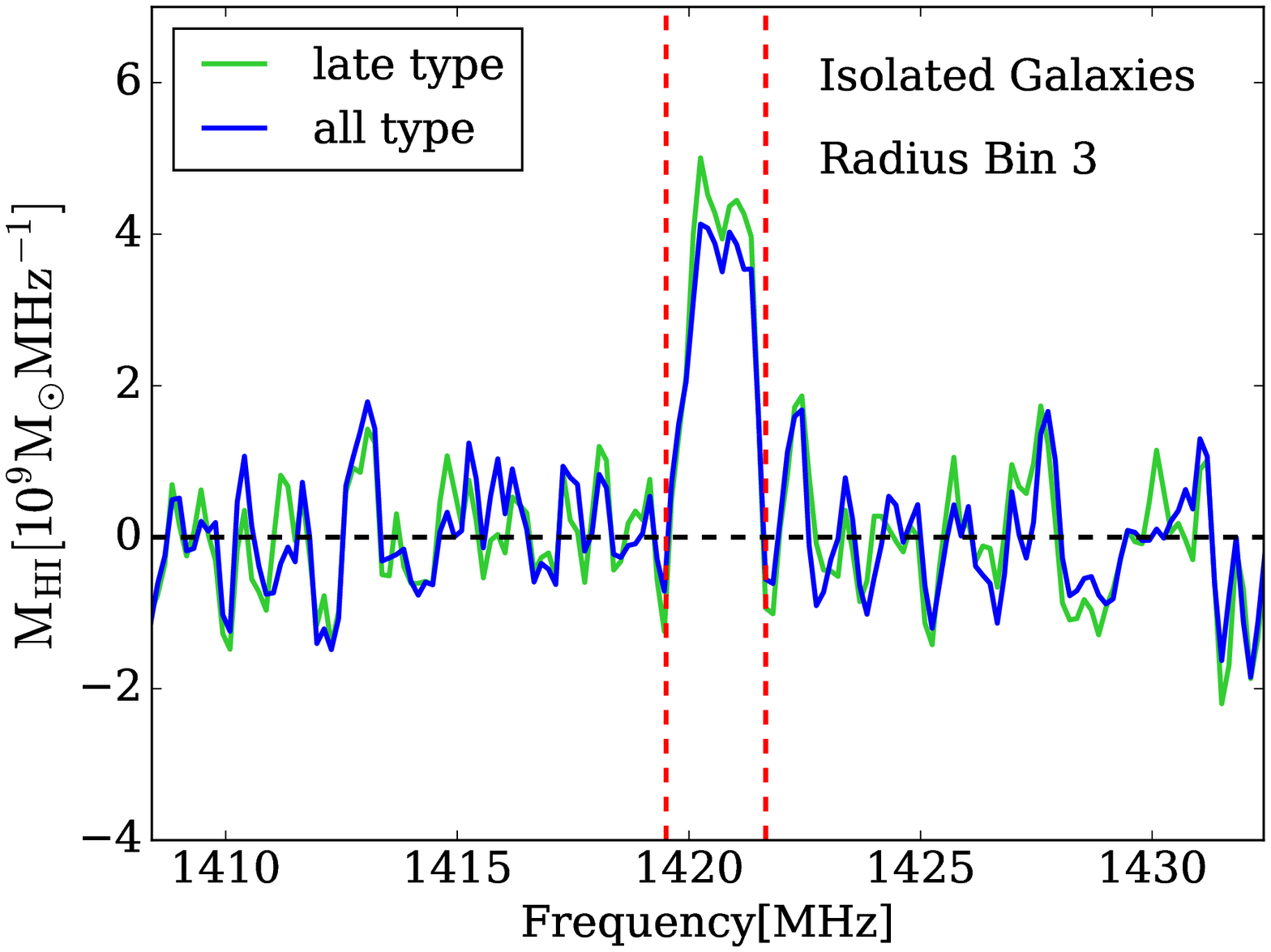}\par
\end{multicols}
\vspace*{-1.05cm}
\begin{multicols}{2}
\includegraphics[width=4.5cm,height=3.75cm]{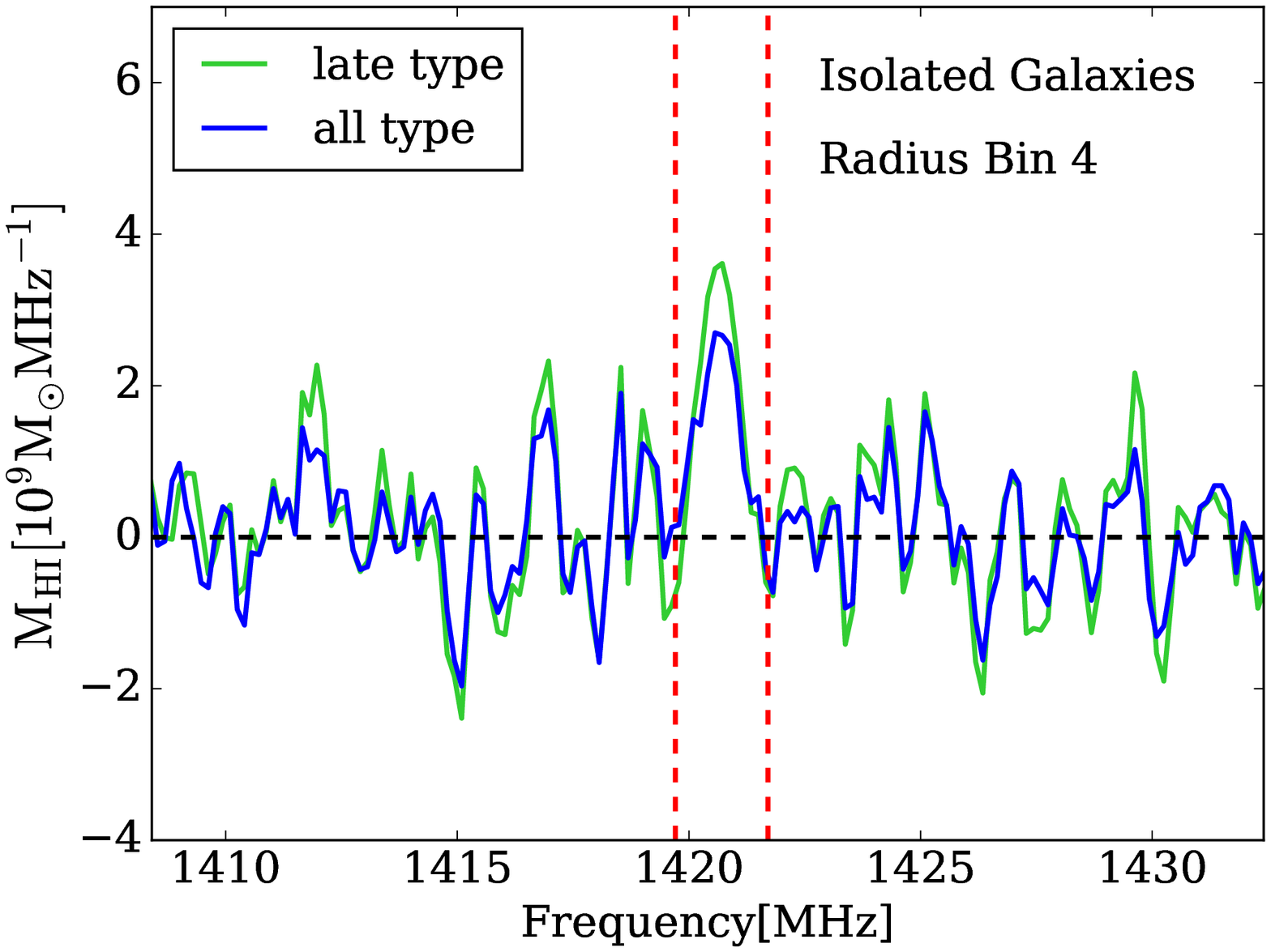}\par
\end{multicols}
\vspace*{-0.8cm}
\caption{The stacked mass spectra for isolated galaxies.}
\label{stacked_spectra_isolate}
\end{figure}

We show the stacked mass spectra for satellite and isolated galaxies in Figure~\ref{stacked_spectra_satellite} and Figure~\ref{stacked_spectra_isolate}. The red-dashed lines show the region over which we do the integration to compute the average \hi\ mass. The spectra for the late-type and the all-type sample are labeled with green and blue, respectively. A larger number of radius bin corresponds to a larger projected radius. Most of stacks give clear detections. But for satellites, the stacks in the first radial bin give marginally detections.

\bsp	
\label{lastpage}
\end{document}